\documentclass[a4paper,12pt]{article} 
\usepackage{axodraw} 

\newcommand{\del}{\partial}

\newcommand{\beq}{\begin{equation}}
\newcommand{\eeq}{\end{equation}}
\newcommand{\bea}{\begin{eqnarray}}
\newcommand{\eea}{\end{eqnarray}}
\newcommand{\bsub}{\begin{subequations}}
\newcommand{\esub}{\end{subequations} \noindent}

\makeatletter
\newtoks\@stequation

\def\subequations{\refstepcounter{equation}%
  \edef\@savedequation{\the\c@equation}%
  \@stequation=\expandafter{\theequation}
  \edef\@savedtheequation{\the\@stequation}
  \edef\oldtheequation{\theequation}%
  \setcounter{equation}{0}%
  \def\theequation{\oldtheequation\alph{equation}}}

\def\endsubequations{%
  \ifnum\c@equation < 2 \@warning{Only \the\c@equation\space subequation
    used in equation \@savedequation}\fi
  \setcounter{equation}{\@savedequation}%
  \@stequation=\expandafter{\@savedtheequation}%
  \edef\theequation{\the\@stequation}%
  \global\@ignoretrue}

\def\eqnarray{\stepcounter{equation}\let\@currentlabel\theequation
\global\@eqnswtrue\m@th
\global\@eqcnt\z@\tabskip\@centering\let\\\@eqncr
$$\halign to\displaywidth\bgroup\@eqnsel\hskip\@centering
     $\displaystyle\tabskip\z@{##}$&\global\@eqcnt\@ne
      \hfil$\;{##}\;$\hfil
     &\global\@eqcnt\tw@ $\displaystyle\tabskip\z@{##}$\hfil
   \tabskip\@centering&\llap{##}\tabskip\z@\cr}

\makeatother

\begin{document} 
\thispagestyle{empty} 
\vspace*{-15mm} 
\baselineskip 10pt 
\begin{flushright} 
\begin{tabular}{l} 
{\bf July 2001}\\ 
{\bf KEK-TH-775}\\ 
{\bf hep-th/0107103} 
\end{tabular} 
\end{flushright} 
\baselineskip 24pt 
\vglue 10mm 
\begin{center} 
{\Large\bf 
 Perturbative world-volume dynamics \\ 
 of the bosonic membrane and string 
}
\vspace{8mm} 

\baselineskip 18pt 
\def\thefootnote{\fnsymbol{footnote}}
\setcounter{footnote}{0}
{\bf 
 Masashi Hayakawa 
  \footnote{e-mail address : haya@post.kek.jp} 
 and 
 Nobuyuki Ishibashi 
  \footnote{e-mail address : ishibash@post.kek.jp} 
} 
\vspace{5mm} 

{\it Theory Division, KEK, Tsukuba, Ibaraki 305-0801, Japan} 

\vspace{10mm} 
\end{center} 

\addtocounter{footnote}{-2} 

\begin{center} 
{\bf Abstract}\\[7mm] 
\begin{minipage}{12cm} 
\baselineskip 16pt 
\noindent 
 We study the world-volume theory of a bosonic membrane 
perturbatively and discuss if one can obtain any conditions 
on the number of space-time dimensions from the consistency 
of the theory. 
We construct an action which is suitable for such a study. 
 In order to study the theory perturbatively
we should specify a classical background 
around which perturbative expansion is defined. 
 We will discuss the conditions which such 
a background should satisfy to deduce the critical dimension. 
 Unfortunately we do not know any background satisfying such 
 conditions. 
 In order to get indirect evidences for the critical dimension 
 of the membrane, 
we next consider two string models obtained via double  
dimensional reduction of the membrane. 
 The first one 
reduces to the Polyakov string theory in the conformal gauge. 
 The second one is described by the Schild action. 
 We show 
that the critical dimension is $26$ for these string theories, 
which implies 
that the critical dimension is $27$ for the membrane theory.

\end{minipage} 
\end{center} 
%
\newpage 
\baselineskip 18pt 
\section{Introduction} 
\label{sec:introduction} 
 Finding a microscopic definition of M-theory \cite{M_theory}  
is one of the most important problems in string theory. 
The low energy effective theory of M-theory is 
the $11$ dimensional supergravity, 
but since it is nonrenormalizable, 
we can do only classical analysis by using it. 
 There exist extended objects, 
membranes and fivebranes in M-theory. 
 Thus it is tempting to consider 
the extended objects in M-theory 
as the fundamental degrees of freedom 
in a microscopic setting of M-theory, 
as was examined in Ref.~\cite{Pioline_01}. 
 It is widely known that the supermembrane 
has an unstable ground state \cite{membrane_deWLN}. 
 In this respect, the M(atrix) theory conjecture \cite{BFSS} 
gives a fascinating explanation which accommodates 
the idea of matrix regularization of the membrane 
\cite{dWHN_membrane}; 
a membrane is a solitonic state 
which can decay into its elementary dynamical constituents. 

 Membranes may not be fundamental objects, 
but it will be possible to get some information about M-theory 
by studying their world-volume quantum dynamics. 
 Although the world-volume theory is nonrenormalizable, 
perturbative analysis as a cut-off theory 
will make it possible for us to discuss 
anomalies in the membrane world-volume gauge symmetries 
in the low energy approximation. 
 The attitude is similar to that of the effective string theory 
\cite{Polchinski_91}. 
 We will restrict ourselves to bosonic membranes in this paper. 
 Such membranes may be related to 
the bosonic string theory via dimensional reduction 
(See Ref.~\cite{bosonic_M} 
for some observation on this point.). 
 Then the critical dimension $D = 27$ is expected to emerge 
as a special number $D$ of target space-time coordinates 
from the requirement of 
the absence of anomalies on the world-volume 
so that its double dimensional reduction naturally yields 
the critical dimension $26$ of the bosonic string theory. 
 This subject has been pursued before in 
a series of papers 
\cite{Marquard_Scholl_1, Marquard_Scholl_2} 
and appearance of some specific features at $D = 27$ 
has been demonstrated. On the other hand, in \cite{Bars:1990ba} 
it was concluded that critical dimension cannot be obtained for 
bosonic membrane theory. 
 See also Ref.~\cite{Deriglazov} for discussion on this topic, 
and Ref.~\cite{Marquard_Sholl_smem,Bars} for supermembranes. 

 In the first part of this paper, we would like to 
reexamine this issue 
by using perturbative analysis 
based on Lagrangian and path integral approach. 
 In order to start perturbative expansions, 
we should specify a classical background of the world-volume theory,
around which the fields fluctuate. 
 We can get conditions on the number of space-time dimensions, 
if the reparametrization symmetry on the world-volume 
becomes anomalous in general. 
 We will argue that, 
in perturbation theory, 
anomaly can occur only 
when there exist no variables 
which can be used as a metric on the world-volume 
being nondegenerate in the classical background. 
 The induced metric $\del_a X^\mu \del_b X_\mu$ 
is such an example. 
 Therefore if the classical value of this is nondegenerate, 
any potential anomaly for the reparametrization symmetry 
can be cancelled. 
 In order to deduce the critical dimension of the theory, 
we should find a classical background in which  
such metrics are singular, and simultaneously the perturbative 
expansion around which is possible. 
 Finding such a background is a difficult problem and 
unfortunately, at present, 
we do not know any examples of classical backgrounds 
that satisfy these conditions. 

 However it is possible to find a classical background 
satisfying such conditions 
after dimensional reduction of the model. 
 The dimensional reduction we will consider is actually 
the double dimensional reduction 
and we get a string model as a result. 
 There are two inequivalent ways of 
double dimensional reduction in 
the world-volume action of membranes we use. 
 One gives the usual Polyakov string theory and the other gives 
a string theory with the Schild action \cite{Schild}.  
 We will demonstrate that the critical dimension 
for such string theories is $26$. 
 This result implies that the critical dimension for 
bosonic membrane theory is $27$. 

 The string theory described by the Schild action 
suffers from a similar difficulty as the one encountered in 
the case of the membrane. 
 In order to analyze the world-sheet theory perturbatively, 
one should specify a classical background and usually 
a good background in this respect yields a nondegenerate 
induced metric. 
 Therefore, it is difficult to deduce the critical dimension 
of such a string theory, 
although we expect that it is $26$. 
 Our result shows how to do so, and may give a clue as to 
how to deduce the critical dimension in the membrane theory. 

 The organization of this paper is as follows.  
 In Sec.~\ref{sec:membrane_dynamics}, we discuss 
bosonic membrane theory. We consider how one can 
deduce the critical dimension of the theory, 
if it is possible. We construct an action 
suitable for perturbative analysis and 
discuss its symmetry. 
 In Sec.~\ref{sec:Polyakov_type} and Sec.~\ref{sec:time_reduction}, 
we discuss two inequivalent ways of dimensional reduction of the 
theory in Sec.~\ref{sec:membrane_dynamics} and 
deduce the critical dimension. 
 Sec.~\ref{sec:discussion} is devoted to discussion. 

\section{Perturbative membrane dynamics} 
\label{sec:membrane_dynamics} 

\subsection{Membrane world-volume action} 
\label{subsec:action} 
In this section, we would like to study the world-volume theory of bosonic membranes 
perturbatively and pursue if there is any critical dimension. Before doing so, 
let us recapitulate the standard procedure of quantizing the membrane theory. 
The action to start from is the Nambu-Goto action:
\begin{equation} 
 S_{NG} 
 = -\int d^3\sigma \sqrt{-\gamma} \, . 
 \label{eq:NG} 
\end{equation} 
 Here $\sigma^a$ ($a=0,1,2$) are the coordinates on the world-volume, 
$X^\mu$ ($\mu$ = 0, 1, $\cdots$, $D-1$) 
represents the embedding of the world-volume 
in the $d$-dimensional space-time, 
$\gamma_{ab}$ = $\del_a X^\mu \del_b X_\mu$ 
is the induced metric and $\gamma = \det(\gamma_{ab})$. 
 Canonical quantization of the action goes in the usual way. 
 The momentum variable $P_\mu$ conjugate to 
$X^\mu$ can be obtained as 
\begin{equation} 
 P_\mu = - \sqrt{-\gamma}\gamma^{0b} \partial_b X_\mu \, . 
\end{equation} 
 One can find the following constraints in the system: 
\begin{eqnarray} 
 &\displaystyle{ 
   \phi_0(\sigma) 
   = \frac{1}{2} \left( 
                   P_\mu(\sigma) P^\mu(\sigma) 
                   + h(\sigma) 
                 \right) \, , 
  }& \nonumber \\ 
 &\displaystyle{ 
  \phi_r(\sigma) 
   = P_\mu(\sigma) \del_r X^\mu(\sigma) \, , 
  }& 
   \label{eq:Poisson_X_P} 
\end{eqnarray} 
with $r = 1, 2$. 
 Here $h(\sigma)$ 
is the determinant 
of the spatial part of the metric $h_{rs}(\sigma)$ 
induced from $X^\mu(\sigma)$, 
\begin{eqnarray} 
 &\displaystyle{ 
   h_{rs}(\sigma) \equiv 
    \del_r X^\mu(\sigma) \del_s X_\mu (\sigma) 
     \, , 
  }& \nonumber \\ 
 &\displaystyle{ 
   h(\sigma) \equiv {\rm det}(h_{rs}(\sigma)) 
    \, . 
  } 
\end{eqnarray} 
 The inverse of $h_{rs}(\sigma)$ will be denoted as 
$h^{rs}(\sigma)$. 
 $\phi_0(\sigma)$ is the Hamiltonian constraint 
and $\phi_r(\sigma)$ are the momentum constraints. 
 They can be regarded as the generators of reparametrization 
on the world-volume. 
 The Hamiltonian made from the action in eq.~(\ref{eq:NG}) 
vanishes as is usual for 
a theory with reparametrization invariance. 

 Using the Poisson bracket 
\begin{equation} 
 \left\{ 
  X^\mu(\sigma^0, \vec{\sigma}), 
  P_\nu(\sigma^0, \vec{\sigma}^{\,\prime}) 
 \right\}_P 
 = \delta^\mu_{\ \nu}\, 
   \delta^2(\vec{\sigma} - \vec{\sigma}^{\,\prime}) \, . 
\end{equation} 
one readily finds 
that all the constraints $\phi_0(\sigma)$, 
$\phi_r(\sigma)$ are of the first class, 
and they satisfy the algebra of 
\begin{eqnarray} 
 && 
  \displaystyle{ 
  \left\{ 
   \phi_0(\vec{\sigma}), \phi_0(\vec{\sigma}^{\,\prime}) 
  \right\}_P 
  = 
  \left( 
   \phi_r(\vec{\sigma}) h(\vec{\sigma}) h^{rs}(\vec{\sigma}) 
   + 
   \phi_r(\vec{\sigma}^{\,\prime}) h(\vec{\sigma}^{\,\prime}) 
    h^{rs}(\vec{\sigma}^{\,\prime}) 
  \right) 
  \del_s 
  \delta^2(\vec{\sigma} - \vec{\sigma}^{\,\prime}) \, , 
  } \nonumber \\ 
 && 
 \displaystyle{ 
  \left\{ 
   \phi_0(\vec{\sigma}), \phi_r(\vec{\sigma}^{\,\prime}) 
  \right\}_P 
  = 
  \left( 
   \phi_0(\vec{\sigma}) + \phi_0(\vec{\sigma}^{\,\prime}) 
  \right) 
  \del_r 
  \delta^2(\vec{\sigma} - \vec{\sigma}^{\,\prime}) \, , 
  } \nonumber \\ 
 && 
 \displaystyle{ 
  \left\{ 
   \phi_r(\vec{\sigma}), \phi_s(\vec{\sigma}^{\,\prime}) 
  \right\}_P 
  = 
    \phi_r(\vec{\sigma}^{\,\prime}) 
     \del_s \delta^2(\vec{\sigma} - \vec{\sigma}^{\,\prime}) 
  + \phi_s(\vec{\sigma}) 
     \del_r \delta^2(\vec{\sigma} - \vec{\sigma}^{\,\prime}) 
      \, . 
  } 
   \label{eq:algebraic_structure} 
\end{eqnarray} 
where $\vec{\sigma} \equiv (\sigma^1, \sigma^2)$, 
and $\sigma^0$-dependence is not made explicit. 
 For a later use, 
we arrange the algebraic structure 
(\ref{eq:algebraic_structure}) in the following form; 
\begin{equation} 
 \left\{ 
  \phi_a(\vec{\sigma}),\, \phi_b(\vec{\sigma}^{\,\prime})  
 \right\}_P 
 = \int d^2\vec{\sigma}^{\,\prime\prime} \, 
    C_{ab}^{\ \ \ c} 
     (\vec{\sigma}, \vec{\sigma}^{\,\prime}; 
      \vec{\sigma}^{\,\prime\prime}) \, 
      \phi_c(\vec{\sigma}^{\,\prime\prime}) \, , 
  \label{eq:constraint_alg} 
\end{equation} 
with a set of $0$-th order structure functions 
$C_{ab}^{\ \ \ c} 
     (\vec{\sigma}, \vec{\sigma}^{\,\prime}; 
      \vec{\sigma}^{\,\prime\prime})$. 
 Their explicit expressions read 
\begin{eqnarray} 
 C_{00}^{\ \ 0}
  (\vec{\sigma}, \vec{\sigma}^{\,\prime}; 
   \vec{\sigma}^{\,\prime\prime}) 
  &=& 0 \, , 
   \nonumber \\ 
 C_{00}^{\ \ r} 
  (\vec{\sigma}, \vec{\sigma}^\prime; 
   \vec{\sigma}^{\,\prime\prime}) 
  &=& 
  \displaystyle{ 
   h(\vec{\sigma}^{\,\prime\prime}) 
    h^{rs}(\vec{\sigma}^{\,\prime\prime}) 
    \del_s \delta^2(\vec{\sigma} - \vec{\sigma}^{\,\prime}) 
    \left( 
     \delta^2(\vec{\sigma} - \vec{\sigma}^{\,\prime\prime}) 
     + 
     \delta^2(\vec{\sigma}^{\,\prime} 
              - \vec{\sigma}^{\,\prime\prime}) 
    \right) \, , 
  } \nonumber \\ 
 C_{0r}^{\ \ 0} 
  (\vec{\sigma}, \vec{\sigma}^{\,\prime}; 
   \vec{\sigma}^{\,\prime\prime}) 
  &=& 
  \displaystyle{ 
   \del_r \delta^2(\vec{\sigma} - \vec{\sigma}^{\,\prime}) 
   \left( 
    \delta^2(\vec{\sigma} - \vec{\sigma}^{\,\prime\prime}) 
    + 
    \delta^2(\vec{\sigma}^{\,\prime} 
             - \vec{\sigma}^{\,\prime\prime}) 
   \right) 
   = C_{r0}^{\ \ 0} 
   (\vec{\sigma}, \vec{\sigma}^{\,\prime}; 
    \vec{\sigma}^{\,\prime\prime}) 
    \, , 
  } \nonumber \\ 
 C_{0r}^{\ \ s} 
  (\vec{\sigma}, \vec{\sigma}^{\,\prime}; 
   \vec{\sigma}^{\,\prime\prime}) 
  &=& 0 
   = C_{r0}^{\ \ s} 
      (\vec{\sigma}, \vec{\sigma}^{\,\prime}; 
       \vec{\sigma}^{\,\prime\prime}) 
        \, , 
   \nonumber \\ 
 C_{rs}^{\ \ 0} 
  (\vec{\sigma}, \vec{\sigma}^{\,\prime}; 
   \vec{\sigma}^{\,\prime\prime}) 
  &=& 0 \, , 
   \nonumber \\ 
  C_{rs}^{\ \ u} 
  (\vec{\sigma}, \vec{\sigma}^{\,\prime}; 
   \vec{\sigma}^{\,\prime\prime}) 
  &=& 
  \displaystyle{ 
   \left( 
    \delta_r^{\ t} \delta_s^{\ u} 
     \delta^2( \vec{\sigma} - \vec{\sigma}^{\,\prime\prime} ) 
    + 
    \delta_r^{\ u} \delta_s^{\ t} 
     \delta^2( \vec{\sigma}^{\,\prime} 
               - \vec{\sigma}^{\,\prime\prime} ) 
   \right) 
   \del_t \delta^2( \vec{\sigma} - \vec{\sigma}^{\,\prime} ) 
    \, . 
  } \label{eq:structure_C} 
\end{eqnarray} 

 If there exists any critical dimension for the membrane theory 
at all, 
Schwinger terms should appear on the right hand side of the above 
algebra when quantized, 
and the condition that they vanish will 
determine the number of the space-time dimensions. 
 There are many ways to quantize the system 
but the critical dimension essentially originates 
from the Schwinger terms in the algebra. 
 The existence of such Schwinger terms implies 
that the diffeomorphism symmetry is anomalous. 
 In the following 
we would like to discuss under what conditions 
such an anomaly can occur. 

 Here we will quantize the system perturbatively. 
 As a three-dimensional field theory, 
the world-volume theory of membranes is nonrenormalizable. 
Hence we consider the theory as a theory with a cut-off. 
The perturbative expansion gives the low energy approximation, 
and we examine if the theory is consistent in the low energy regime. 
 The action most convenient for the perturbative analysis of this system 
can be obtained as follows. 
 Since there are constraints, 
we introduce the Lagrangian multiplier fields 
$\lambda^0(\sigma)$, $\lambda^r(\sigma)$ ($r = 1, 2$) to 
respective constraints and we have the Hamiltonian:
\begin{eqnarray} 
 H &=& 
 \displaystyle{ 
  \int d^2 \vec{\sigma}\, 
   \left( 
      \lambda^0(\sigma) \phi_0(\sigma) 
    + \lambda^r(\sigma) \phi_r(\sigma) 
   \right) 
 } \nonumber \\ 
 &=& 
 \displaystyle{ 
  \int d^2 \vec{\sigma}\, 
   \left[ 
    \lambda^0(\sigma) \, 
    \frac{1}{2} 
    \left( 
       P_\mu(\sigma) P^\mu(\sigma) 
     + h(\sigma) 
    \right) 
    + 
    \lambda^r(\sigma) 
    P_\mu(\sigma) \del_r X^\mu(\sigma) 
   \right] \, . 
 } 
  \label{eq:Hamiltonian} 
\end{eqnarray} 
 By Legendre transformation, we obtain the action 
\begin{eqnarray} 
 S_0 &=& 
 \displaystyle{ 
    \int d^3 \sigma \, P_\mu(\sigma) \del_0 X^\mu(\sigma) 
  - \int d \sigma^0\, H 
 } \nonumber \\ 
 &=& 
 \displaystyle{ 
  \int d^3 \sigma \, 
  \left( 
    P_\mu(\sigma) \del_0 X^\mu(\sigma) 
  \right. 
 } \nonumber \\ 
 && \qquad \quad 
 \displaystyle{ 
  \left. 
    - 
    \lambda^0(\sigma) \, 
    \frac{1}{2} 
    \left( 
       P_\mu(\sigma) P^\mu(\sigma) 
     + h(\sigma) 
    \right) 
    - 
    \lambda^r(\sigma) 
    P_\mu(\sigma) \del_r X^\mu(\sigma) 
  \right) \, . 
 } \label{eq:action_def} 
\end{eqnarray} 
 Gaussian integration of (\ref{eq:action_def}) 
over momenta $P_\mu(\sigma)$ gives 
\begin{eqnarray} 
 S_0 
 &=& 
 \displaystyle{ 
  \int d^3 \sigma 
  \left( 
   \frac{1}{2\lambda^0} \, 
   \left( 
    \del_0 X^\mu - \lambda^r \del_r X^\mu 
   \right)^2 
   - 
   \frac{1}{2} \lambda^0 h 
  \right) 
 } \nonumber \\ 
 &=& 
 \displaystyle{ 
  \int d^3 \sigma 
  \left( 
   \frac{1}{2\lambda^0} \, 
   \left( 
    \del_0 X^\mu - \lambda^r \del_r X^\mu 
   \right)^2 
   - 
   \frac{\lambda^0}{4}\, 
   \left( 
    \left\{ 
     X^\mu, X^\nu 
    \right\} 
   \right)^2 
  \right) \, , 
 } 
 \label{eq:action_main}
\end{eqnarray} 
where 
\begin{equation} 
 \left\{ A, B \right\} 
  \equiv \epsilon^{rs} \del_r A \del_s B \, . 
\end{equation} 
 This is the action we start from. 
 This action was obtained in \cite{Barcelos}. 
 Since it is in the form of polynomials 
of $\partial_a X^\mu$, 
after an appropriate gauge fixing and expansion around 
an appropriate background, 
we will be able to get an action in which perturbative analysis 
is possible. 

\subsection{BRS transformation} 
\label{subsec:BRS} 

 Since we would like to study if the symmetry corresponding to 
the constraints $\phi_0$, $\phi_r$ is anomalous or not 
by using the action (\ref{eq:action_main}), 
we need to know how such a symmetry is realized 
in the action (\ref{eq:action_main}). 
 In this subsection, we will discuss the local symmetry of the action 
and treat it by using the BRS formalism. 

 The symmetry 
generated by $\phi_0$, $\phi_r$ is realized in 
eq.~(\ref{eq:action_main}) as 
\begin{eqnarray} 
 \delta_\epsilon X^\mu 
 &=& 
 \displaystyle{ 
  \epsilon^0 \frac{1}{\lambda^0} 
  \left( 
   \del_0 X^\mu - \lambda^r \del_r X^\mu 
  \right) 
  + 
  \epsilon^r \del_r X^\mu \, , 
 } \nonumber \\ 
 \delta_\epsilon \lambda^0 
 &=& 
 \displaystyle{ 
  \del_0 \epsilon^0 
  + \epsilon^0 \del_r \lambda^r 
  - \del_r \epsilon^0 \lambda^r 
  + \epsilon^r \del_r \lambda^0 
  - \del_r \epsilon^r \lambda^0 \, , 
 } \nonumber \\ 
 \delta_\epsilon \lambda^r 
 &=& 
 \displaystyle{ 
  \del_0 \epsilon^r 
  + 
  h h^{rs} 
  \left( 
     \epsilon^0 \del_s \lambda^0 
   - \del_s \epsilon^0 \lambda^0 
  \right) 
  - 
  \lambda^s \del_s \epsilon^r 
  + 
  \epsilon^s \del_s \lambda^r \, , 
 } \label{eq:gauge_tr_original} 
\end{eqnarray} 
with three parameters, $\epsilon^0$, 
$\epsilon^r$ ($r = 1, 2$). 
 In eq.~(\ref{eq:gauge_tr_original}), 
the transformation law for $X^\mu$ has been inferred from 
that in the Hamiltonian description 
\begin{eqnarray} 
 \delta_\epsilon X^\mu(\vec{\sigma}) 
 &=& 
 \displaystyle{ 
  \int d^2 \vec{\sigma}^{\,\prime}\, 
  \left\{ 
    X^\mu(\vec{\sigma}), \, 
     \phi_a(\vec{\sigma}^{\,\prime}) 
     \epsilon^a(\vec{\sigma}^{\,\prime}) 
  \right\}_P 
 } \nonumber \\ 
 &=& 
 \displaystyle{ 
    \epsilon^0(\vec{\sigma}) P^\mu(\vec{\sigma}) 
  + \epsilon^r(\vec{\sigma}) \del_r X^\mu(\vec{\sigma}) 
     \, , 
 } 
\end{eqnarray} 
with the use of the value of $P_\mu$ obtained from 
the equations of motion of the action (\ref{eq:action_def}) 
\begin{equation} 
 P^\mu 
 = 
 \frac{1}{\lambda^0} 
 \left( 
  \del_0 X^\mu - \lambda^r \del_r X^\mu 
 \right) \, . 
\end{equation} 
 In general, 
the action becomes invariant 
if the Lagrange multipliers transform in the manner 
\begin{equation} 
 \delta_\epsilon \lambda^a(\vec{\sigma}) 
 = \del_0 \epsilon^a(\vec{\sigma}) 
   - \int d^2 \vec{\sigma}^{\,\prime} 
     \int d^2 \vec{\sigma}^{\,\prime\prime} 
     \epsilon^c(\vec{\sigma}^{\,\prime\prime}) 
     \lambda^b(\vec{\sigma}^{\,\prime}) 
     C_{bc}^{\ \ a} 
      (\vec{\sigma}^{\,\prime}, 
       \vec{\sigma}^{\,\prime\prime}, 
       \vec{\sigma}) \, , 
\end{equation} 
using the structure functions 
$C_{ab}^{\ \ \ c} 
     (\vec{\sigma}, \vec{\sigma}^{\,\prime}; 
      \vec{\sigma}^{\,\prime\prime})$ 
of the constraint algebra (\ref{eq:constraint_alg}). 
 The explicit form of $C_{ab}^{\ \ \ c} 
     (\vec{\sigma}, \vec{\sigma}^{\,\prime}; 
      \vec{\sigma}^{\,\prime\prime})$ 
in eq.~(\ref{eq:structure_C}) led to 
the last two equations in eq.~(\ref{eq:gauge_tr_original}). 
 Defining $\widehat{\epsilon}^{\,0}$, 
$\widehat{\epsilon}^{\,r}$ as  
\begin{equation} 
 \widehat{\epsilon}^{\,0} \equiv 
  \epsilon^0 \frac{1}{\lambda^0} \, , \quad 
 \widehat{\epsilon}^{\,r} \equiv 
  \epsilon^r - \widehat{\epsilon}^{\,0} \lambda^r \, , 
\end{equation} 
the gauge transformation (\ref{eq:gauge_tr_original}) 
becomes 
\begin{eqnarray} 
 \delta_\epsilon X^\mu 
 &=& 
 \displaystyle{ 
  \widehat{\epsilon}^{\,a} \del_a X^\mu \, , 
 } \nonumber \\ 
 \delta_\epsilon \lambda^0 
 &=& 
 \displaystyle{ 
  \widehat{\epsilon}^{\,a} \del_a \lambda^0 
  + 
  \lambda^0 
  \left( 
   \del_0 \widehat{\epsilon}^{\,0} 
   - 
   \del_r \widehat{\epsilon}^{\,r} 
   - 
   2 \lambda^r \del_r \widehat{\epsilon}^{\,0} 
  \right) 
   \, , 
 } \nonumber \\ 
 \delta_\epsilon \lambda^r 
 &=& 
 \displaystyle{ 
  \widehat{\epsilon}^{\,a} \del_a \lambda^r 
  + 
  \left( 
   \del_0 - \lambda^s \del_s 
  \right) 
    \widehat{\epsilon}^{\,r} 
  + 
  \lambda^r \del_0 \widehat{\epsilon}^{\,0} 
  - 
  I^{rs} 
   \del_s \widehat{\epsilon}^{\,0} 
    \, , 
 } \label{eq:gauge_transformation} 
\end{eqnarray} 
where 
\begin{equation} 
 I^{rs} 
 \equiv 
 (\lambda^0)^2 h h^{rs} 
 + 
 \lambda^r \lambda^s 
  \, . 
\end{equation} 
 The transformation law for $X^\mu$ 
in eq.~(\ref{eq:gauge_transformation}) 
shows that 
the gauge symmetry 
corresponds to the reparametrization on the world-volume. 

 In order to treat this theory with local symmetries perturbatively, 
we will use the BRS formalism. 
 Since the structure constants of the symmetry are field-dependent, 
we should resort to the Batalin-Vilkoviski formalism\cite{Batalin:1981jr, HT}. 
 In order to do so,  
let us explore the algebraic structure 
of the gauge transformation (\ref{eq:gauge_transformation}).
\footnote{We will follow the notation used in \cite{HT} in the following.} 
 The gauge transformation is summarized abstractly by 
\begin{equation} 
 \delta_\epsilon \phi^i = \epsilon^\alpha R_\alpha^{\ i} 
  \, , 
   \label{eq:gauge_abstract} 
\end{equation} 
where 
the index $i$ distinguishes the dynamical variables 
while $\alpha$ labels the 
gauge degrees of freedom. 
 We adopt the convention that 
both $i$ and $\alpha$ include the dependence 
on the world-volume coordinates. 
 The algebraic structure possessed by 
the gauge symmetry (\ref{eq:gauge_abstract}) 
is summarized as 
\begin{equation} 
 R_\alpha^{\ j} 
  \frac{\delta R_\beta^{\ i}}{\delta \phi^j} 
 - 
 R_\beta^{\ j} 
  \frac{\delta R_\alpha^{\ i}}{\delta \phi^j} 
 = 
 D_{\alpha \beta}^{\ \ \ \gamma} R_\gamma^{\ i} 
 + 
 M_{\alpha\beta}^{\ \ \ ij} 
  \frac{\delta S_0}{\delta \phi^j} 
   \, . 
    \label{eq:algebra_general} 
\end{equation} 
 Here $D_{\alpha\beta}^{\ \ \gamma}$ are 
the structure functions which can be read off as 
\begin{eqnarray} 
 && 
 D_{ab}^{\ \ c}(\sigma^\prime, \sigma^{\prime\prime}; \sigma) 
  \nonumber \\ 
 && \qquad 
 = 
  \delta_a^{\ c} 
   \del_b \delta^3(\sigma - \sigma^\prime) 
   \delta^3(\sigma - \sigma^{\prime\prime}) 
  - 
  \delta_b^{\ c} 
   \delta^3(\sigma - \sigma^\prime) 
   \del_a \delta^3(\sigma - \sigma^{\prime\prime}) \, . 
\end{eqnarray} 
 The second term in eq.~(\ref{eq:algebra_general}) 
corresponds to the trivial symmetry 
\begin{equation} 
 \delta_{\mu} \phi^j 
 = 
 \mu^{jk} \frac{\delta S_0}{\delta \phi^k} \, , 
\end{equation} 
with the parameters $\mu^{jk} = - \mu^{kj}$. 
 A straightforward calculation shows that 
\begin{eqnarray} 
 && 
 M_{ab}^{\ \ ru} 
  \left( 
   \sigma^\prime, \sigma^{\prime\prime}; 
   \sigma, \sigma^{\prime\prime\prime} 
  \right) 
   \nonumber \\ 
 && \qquad 
 = 
 3 [\lambda^0(\sigma)]^3 \, 
 \delta_a^{\ 0} \delta_b^{\ 0} \, 
 \epsilon^{st} \, 
  \del_t \delta^3(\sigma - \sigma^\prime) 
  \del_s \delta^3(\sigma - \sigma^{\prime\prime}) \, 
  \epsilon^{ru} 
  \delta^3(\sigma - \sigma^{\prime\prime\prime}) \, . 
\end{eqnarray} 
 With these ingredients we can 
apply the Batalin-Vilkoviski procedure 
to the membrane quantum mechanics 
and obtain a solution of the classical master equation 
as well as the BRS transformation. 

 A solution of the classical master equation is found as 
\begin{eqnarray} 
 && 
 S = ^{(0)}S + ^{(1)}S + ^{(2)}S \, , 
  \nonumber \\ 
 && 
 ^{(0)}S = S_0 = 
   \int d^3 \sigma 
    \left( 
     \frac{1}{\lambda^0} \frac{1}{2} 
      \left( 
       \dot{X}^\mu - \lambda^r \del_r X^\mu 
      \right)^2 
      - 
      \frac{1}{2} 
      \lambda^0 
      h 
    \right) \, , 
     \nonumber \\ 
 && 
 ^{(1)}S = 
   \int d^3 \sigma 
   \left( 
    X_\mu^* \del_a X^\mu C^a 
   \right. 
    \nonumber \\ 
 && \qquad \qquad 
    - \dot{\lambda^*_0} \lambda^0 C^0 
    - 2 \lambda^*_0 \lambda^0 \lambda^r \del_r C^0 
    + \lambda^*_0 
       \left( 
        -\lambda^0 \del_r + (\del_r \lambda^0) 
       \right) C^r 
    \nonumber \\ 
 && \qquad \qquad 
   \left. 
    - \dot{\lambda^*_r} \lambda^r C^0 
    - \lambda^*_r I^{rs} \del_s C^0 
    + \lambda^*_r 
       \left( 
        \del_0 - \lambda^s \del_s 
       \right) C^r 
    + \lambda^*_r (\del_s \lambda^r) C^s 
   \right) \, , 
    \nonumber \\ 
 && 
 ^{(2)}S = 
 \int d^3 \sigma 
 \left( 
  -C^*_a C^b \del_b C^a 
  + 
  \frac{3}{4} \, 
   (\lambda^0)^3
   \epsilon^{ru} \lambda^*_r \lambda^*_u \, 
   \left\{ C^0, C^0 \right\} 
 \right) \, , 
  \label{eq:BV_membrane_action} 
\end{eqnarray} 
 Here, 
$X^*_\mu$, $\lambda^*_a$ and $C^*_a$ 
are the antifields of 
$X^\mu$, $\lambda^a$ and $C^a$ respectively. 
 The number $i$ in the parenthesis of $^{(i)} S$ 
denotes the antighost number. 
 The BRS transformation is generated 
by $S$ in terms of the antibracket, 
\begin{eqnarray} 
 && 
 s \phi^i = (\phi_i, S) 
  = \frac{\delta S}{\delta \phi^*_i} \, , 
  \nonumber \\ 
 && 
 s \phi^*_i = (\phi^*_i, S) 
  = - \frac{\delta S}{\delta \phi^i} \, , 
\end{eqnarray} 
where the functional derivatives are 
understood to act from the left. 
 The explicit expression for the BRS transformation law 
becomes, for instance, 
\begin{eqnarray} 
 && 
 s X^\mu = \del_a X^\mu C^a \, , 
  \nonumber \\ 
 && 
 s \lambda^0 
 = 
   \del_0 (\lambda^0 C^0) 
 - 2 \lambda^0 \lambda^r \del_r C^0 
 - 
 \left( 
  \lambda^0 \del_r - (\del_r \lambda^0) 
 \right) C^r 
  \, , 
   \nonumber \\ 
 && 
 \displaystyle{ 
  s \lambda^r 
  = 
    C^a \del_a \lambda^r 
  + \lambda^r \del_0 C^0 
  - I^{rs} \del_s C^0 
  + \left( 
     \del_0 - \lambda^s \del_s 
    \right) C^r 
  + \frac{3}{2} (\lambda^0)^3 
    \epsilon^{ru} 
    \lambda^*_u \left\{ C^0, C^0 \right\} 
     \, , 
 } \nonumber \\ 
 && 
 s \lambda^*_r 
 = 
 \frac{1}{\lambda_0} 
  \left( 
     \del_0 X_\mu \del_r X^\mu 
   - \lambda^s h_{sr} 
  \right) 
    \nonumber \\ 
 && \qquad \qquad 
 + 2 \lambda^0 \lambda^*_0 \del_r C^0 
 + \del_0 \lambda^*_r C^0 
 + \lambda^*_r \lambda^s \del_s C^0 
 + \lambda^*_s \lambda^s \del_r C^0 
    \nonumber \\ 
 && \qquad \qquad 
 + \lambda^*_s \del_r C^s 
 + \del_s (\lambda^*_r C^s) 
  \, , 
   \nonumber \\ 
 && 
 s C^a 
 = C^b \del_b C^a 
  \, . 
   \label{eq:BRS} 
\end{eqnarray} 

\subsection{BRS covariant metric and perturbative anomaly} 
\label{subsec:BRS_metric} 
 Now let us consider the question 
under what conditions the reparametrization symmetry 
or the BRS symmetry of the action $S$ 
in eq.~(\ref{eq:BV_membrane_action}) can become anomalous. 
 In order to do so, 
it is instructive to recall how the reparametrization symmetry  
becomes anomalous in the case of string theory. 
 In the string case, 
starting from the Nambu-Goto action, we can follow the same 
procedure as that in Sec.~\ref{subsec:action} 
and obtain the action 
\beq 
 S_0 = 
  \int d^2 \sigma 
  \left( 
   \frac{1}{2\lambda^0} \, 
   \left( 
    \del_0 X^\mu - \lambda^1 \del_1 X^\mu 
   \right)^2 
   - 
   \frac{\lambda^0}{2}\, 
   \left(  
     \del_1X^\mu 
   \right)^2 
  \right) \, . 
 \label{barcelosstring} 
\eeq 
 This action is actually 
the Polyakov string action \cite{Polyakov}
\beq 
 \int d^2 \sigma \, 
 \frac{1}{2} \sqrt{-g} g^{ab} \del_a X^\mu \del_b X_\mu \, , 
\eeq 
with $\sqrt{-g} g^{ab}$ ($a, b = 0, 1$) expressed in terms of 
$\lambda^0$, $\lambda^1$. 
 Since $\sqrt{-g}g^{ab}$ do not include 
the conformal mode of the metric, 
$\lambda^0$ and $\lambda^1$ 
express the modes of the metric other 
than the conformal mode. 

 Starting from the action in eq.~(\ref{barcelosstring}) 
one can quantize the theory 
using the BRS formalism in the usual way and find that the BRS symmetry 
is anomalous 
if $D\neq 26$. 
 The reason for the existence of the anomaly is obvious. 
 In order to define a quantum theory, 
we need a metric to define the regularization 
procedure, etc.~\cite{Fujikawa:1983im}. 
 If a metric $g_{ab}$ which is transformed as a tensor 
under the reparametrization is available, 
we can regularize the action 
by using the Laplacian made from  $g_{ab}$ for example. 
Then the theory can be defined preserving 
the reparametrization symmetry. 
In order to regularize the action in the BRS invariant manner, 
we need a metric $g_{ab}$ which is transformed 
under the BRS symmetry as 
\begin{equation} 
 s g_{ab} 
  = g_{ac} \del_b C^c + \del_a C^c g_{cb} 
     + C^c \del_c g_{ab} \, , 
\end{equation} 
where $C^a$ ($a = 0, 1$) are the reparametrization ghost fields. 
Let us call such a metric a {\it BRS covariant metric}. 
 If a metric which behaves properly under 
the reparametrization symmetry or BRS symmetry is not available, 
it can be a source of an anomaly. 
 In the present case, we have the variables $\lambda^0$, $\lambda^1$ 
from which we can construct $\sqrt{-g}g^{ab}$ 
but not $g_{ab}$ itself. 
 Therefore we cannot have a metric 
which behaves properly under the symmetry, 
and the symmetry becomes anomalous. 
 On the other hand, in the case of Polyakov string theory, 
we consider $g_{ab}$ as the fundamental degrees of freedom and 
the reparametrization invariance can be made nonanomalous. 
 Then what matters is the Weyl symmetry, 
or the conformal symmetry in the conformal gauge. 

 Actually in the formulation 
using the action in eq.~(\ref{barcelosstring}), 
things are more subtle than it appears. 
 It is not possible to construct a metric 
from $\lambda^0$, $\lambda^1$ alone, 
but it is possible to do so using $X^\mu$. 
 Indeed, the induced metric $\del_a X^\mu \del_b X_\mu$ 
is a BRS covariant metric. 
 The reason why such a metric is not considered 
to make the symmetry nonanomalous is 
because we always consider the world-sheet theory 
of strings around the classical background $X^\mu =0$. 
 Therefore, at least perturbatively, 
the induced metric is a singular metric and cannot be used. 

 In Ref.~\cite{Polchinski_91}, 
Polchinski and Strominger studied Nambu-Goto string theory 
perturbatively around a background. 
 For a perturbation theory to be well-defined, 
the induced metric, 
necessarily becomes nondegenerate. 
 Hence, an anomaly in the reparametrization symmetry  
can be cancelled in this setting. 
Indeed they showed  
that it is possible to construct the counterterms 
which cancel the potential anomaly 
for the reparametrization invariance in such a background. 

 Now let us turn to the membrane theory. 
 The action in eq.~(\ref{eq:action_main}) can be considered 
as a generalization of the action in eq.~(\ref{barcelosstring}), 
but there are several differences. 
 Firstly, this action does not coincide 
with the Polyakov-type action with an intrinsic metric $g_{ab}$, 
but rather can be obtained 
by classically integrating out the spatial metric 
in the Polyakov-type action 
after taking a gauge considered in \cite{Fujikawa_Kubo}. 
 Since the Polyakov-type action 
does not possess a symmetry like the conformal symmetry, 
we cannot attribute 
the problem to such a symmetry being anomalous. 
 Another difference is 
that we cannot obtain a well-defined perturbation theory from 
the action in eq.~(\ref{eq:action_main}) 
by expanding around the classical background $X^\mu =0$. 
 Therefore we should specify a good classical background 
to start perturbative expansions. 

 The problem is if a metric which behaves properly 
under the symmetry is available or not 
in the classical background. 
 Also in this case, 
one of such metrics is the three-dimensional induced metric 
\begin{equation} 
 \gamma_{ab} = \del_a X^\mu \del_b X_\mu 
  \quad (a, b = 0, 1, 2) \, . 
  \label{eq:metric_gamma} 
\end{equation} 
 We can construct another BRS covariant metric $g_{ab}$ 
defined as  
\begin{eqnarray} 
 g_{ab} 
 &=& 
 \displaystyle{ 
  \left[ 
   \begin{array}{cc} 
    \lambda^t G_{tu} \lambda^u - (\lambda^0)^2 G & 
    G_{st} \lambda^t \\ 
    G_{rt} \lambda^t & G_{rs} 
   \end{array} 
  \right] \, , 
 } \nonumber \\ 
 g^{ab} 
 &=& 
 \displaystyle{ 
  \left[ 
   \begin{array}{cc} 
    - \frac{1}{(\lambda^0)^2 G} & 
    \frac{\lambda^s}{(\lambda^0)^2 G} \\ 
    \frac{\lambda^r}{(\lambda^0)^2 G} & 
    G^{rs} - \frac{\lambda^r \lambda^s}{(\lambda^0)^2 G} 
   \end{array} 
  \right] \, . 
 } 
  \label{eq:three_metric} 
\end{eqnarray} 
 Here 
\begin{equation} 
 G_{rs} \equiv \del_rX^\mu\del_sX_\mu 
  + \lambda^0 
    \left( 
     \lambda^*_r \del_s C^0 + \lambda^*_s \del_r C^0 
    \right) 
  \, .
   \label{eq:def_of_G} 
\end{equation} 
 The metric in eq.~(\ref{eq:three_metric}) is in the form 
where $\sqrt{G}\lambda^0$ is the lapse, $\lambda^r$ are the shift 
and $G_{rs}$ is the space metric in the ADM decomposition 
of the metric \cite{Gravitation}. 
 $G_{rs}$ is exactly 
the combination appeared in \cite{Fujikawa_Kubo}. 
 A similar kind of metric 
can also be constructed in the same fashion 
for higher dimensional branes and strings. 

 It is of course probable 
that there exist other BRS covariant metrics. 
 The potential anomaly of the BRS symmetry 
can be cancelled by adding counterterms, 
if any one of such metrics is nondegenerate 
in the classical background. 
 The classical background we should consider is 
a solution of the classical equations of motion 
around which the fields corresponding to 
quantum fluctuations obtain 
regular propagators in all directions 
after a proper gauge fixing. 

 One typical solution to realize this last requirement 
is an infinitely extended static membrane configuration; 
\begin{eqnarray} 
 &X^0_{(0)}(\sigma) = \sigma^0 \, , \quad 
  X^1_{(0)}(\sigma) = \sigma^1 \, , \quad  
  X^2_{(0)}(\sigma) = \sigma^2 \, ,& 
   \nonumber \\ 
 &X^I_{(0)}(\sigma) = 0 \quad (I=3, \cdots, D-1) \, ,& 
  \label{eq:classical_bg} 
\end{eqnarray} 
with $\lambda^0_{(0)}(\sigma) = 1$ 
and $\lambda^r_{(0)}(\sigma) = 0$. 
 We introduce a coupling constant $g$ 
characterizing the order of perturbation. 
 Accordingly, 
the left hand side of eq.~(\ref{eq:BV_membrane_action}) 
is regarded as $g^2 S$ and 
all the fields $\phi$ except for $X^\mu$ 
should be replaced by $g \phi$. 
 For the choice of the gauge fixing condition 
\begin{eqnarray} 
 & 
 \lambda^0(\sigma) = 1 \, ,& 
  \nonumber \\ 
 & 
  \lambda^r(\sigma) = g\, \epsilon^{rs} \del_s A_0(\sigma) \, ,& 
\end{eqnarray} 
the fluctuations around the background (\ref{eq:classical_bg}), 
\begin{eqnarray} 
 & 
  X^r(\sigma) = \sigma^r + g \epsilon^{rs} A_s(\sigma) 
  \quad (r = 1, 2) \, ,& 
   \nonumber \\ 
 & 
  X^A(\sigma) = g Y^A(\sigma) 
   \quad (A = 0, 3, \cdots, D-1) \, ,& 
\end{eqnarray} 
are governed by the action similar to the Yang-Mills action 
\begin{equation} 
 S_0 
 = 
   \int d^3 \sigma \, 
   \left( 
     \frac{1}{g^2} 
     - 
     \frac{1}{4} F_{ab} F^{ab} 
     - 
     \frac{1}{2} D_a Y^A D^a Y^A 
     - 
     \frac{g^2}{4} 
      \left( 
       \left\{ Y^A, Y^B \right\} 
      \right)^2 
   \right) \, , 
\end{equation} 
where 
\begin{eqnarray} 
 && 
  F_{ab} \equiv 
   \del_a A_b - \del_b A_a 
    + g \left\{ A_a, A_b \right\} \, , 
     \nonumber \\ 
 && 
 D_a Y^A \equiv 
  \del_a Y^A 
  + g \left\{ A_a, Y^A \right\} \, . 
\end{eqnarray} 
 After setting a gauge fixing condition 
as in the usual Yang-Mills theory, 
all the fields have free propagators. 
 However, the metrics (\ref{eq:metric_gamma}) and 
(\ref{eq:three_metric}) 
are both regular in the background (\ref{eq:classical_bg}). 
 Thus, a path integral measure invariant under the BRS symmetry 
can be constructed using either of these metrics 
so that the perturbative quantum theory 
around (\ref{eq:classical_bg}) does not have anomalies. 
 Although we do not dictate its detail, 
it is tedious but straightforward to perform 
the explicit one-loop analysis similar to that 
to be performed  
in Sec.~\ref{subsec:anomaly_bosonic_Schild} 
and show the absence of an anomaly 
in the perturbation theory around the static membrane 
solution (\ref{eq:classical_bg}). 

 Obviously the discussion here applies to any backgrounds. 
 An anomaly can arise only on the background which 
gives no regular BRS covariant metrics.  
 Unfortunately, at present we do not know 
any example of such a background 
around which all the fields have propagator suitable 
for perturbation theory. 
 However it is possible 
to find a background around which the kinetic terms 
for some fields lack derivatives 
in some directions on the world-volume. 
In such a case, 
we encounter severe divergences in perturbative expansions, 
because propagators in the position space 
include delta function of coordinates in some directions. 
 However, even in such a background, 
if one dimensionally reduces the theory, one can get a 
well-defined perturbation theory. 
 Then we can get a little indirect evidence 
for the critical dimension of membrane theory. 
 We will consider two examples 
in Sec.~\ref{sec:Polyakov_type} 
and Sec.~\ref{sec:time_reduction}. 

\section{Dimensional reduction of one spatial direction} 
\label{sec:Polyakov_type} 

 Let us consider the following 
configuration of the membrane world-volume 
\begin{eqnarray} 
 &X^2_{(0)}(\sigma) = \sigma^2 \, ,& 
   \nonumber \\ 
 &X^M_{(0)}(\sigma) = 0 
  \quad (M = 0, 1, 3, \cdots, D-1) \, .& 
   \label{eq:classical_bg_2} 
\end{eqnarray} 
 It is easy to check that this is a solution of 
the equation of motion 
of the action (\ref{eq:BV_membrane_action}). 
 The fluctuation $\widehat{Y}^\mu$ defined by 
\begin{eqnarray} 
 X^2 &=& \sigma^2 + g \widehat{Y}^2 \, , 
  \nonumber \\ 
 X^M &=& g \widehat{Y}^M \ (M = 0, 1, 3, \cdots, D-1) \, , 
\end{eqnarray} 
is governed by the action 
\begin{eqnarray} 
 S_0 &=& 
 \displaystyle{ 
  \int d^3 \sigma\, 
  \left[ 
     \frac{1}{2 \lambda^0} 
     \left( 
      - \frac{\lambda^2}{g} 
      + \del_0 \widehat{Y}^2 - \lambda^r \del_r \widehat{Y}^2 
     \right)^2 
   + \frac{1}{2 \lambda^0} 
     \left( 
      \del_0 \widehat{Y}^M - \lambda^r \del_r \widehat{Y}^M 
     \right)^2 
  \right. 
 } \nonumber \\ 
 && \qquad \quad 
 \displaystyle{
  \left.
  - \frac{\lambda^0}{2} (\del_1 \widehat{Y}^M)^2 
  - g^2 \frac{\lambda^0}{4} 
     \left( 
      \left\{ \widehat{Y}^\mu, \widehat{Y}^\nu \right\} 
     \right)^2 
  \right]
     \, .
 } \label{eq:action_before} 
\end{eqnarray} 
 Taking the gauge $\lambda^0=1,~\lambda^r=0$, 
the kinetic terms for $\widehat{Y}^M$ lack the 
derivatives in $\sigma^2$ direction. 
 Thus, we will encounter severe divergences in the 
perturbative expansion.  

 In order to avoid this problem, 
let us compactify the $X^2$-direction 
to a circle and reduce its radius $R$ to zero.
 The background (\ref{eq:classical_bg_2}) implies that 
we are considering the membrane wrapped around $X^2$-direction. 
 Therefore, 
what we are doing is the so-called double dimensional 
reduction and as a result we get a string theory. 
 Since the action we started from was a generalization of 
the Polyakov string action, 
we expect that we get the Polyakov string action 
after the dimensional reduction. We will demonstrate that 
this is indeed the case. 

 In the limit $R\rightarrow 0$, we will keep 
the coupling constant $g_2 \equiv g/ \sqrt{2\pi R}$ 
 fixed. 
 The fluctuation in the world-sheet theory 
corresponds to 
$Y^\mu \equiv \sqrt{2 \pi R} \widehat{Y}^\mu$. 
 Redefining $g_2$ as new $g$ and dropping 
all the derivatives in $\sigma^2$-direction, 
eq.~(\ref{eq:action_before}) reduces to a 
two-dimensional theory 
\begin{eqnarray} 
 S_0 &=& 
 \displaystyle{ 
  \int d^2 \sigma\, 
  \left[ 
     \frac{1}{2 \lambda^0} 
     \left( 
      - \frac{\lambda^2}{g} 
      + \del_0 Y^2 - \lambda^1 \del_1 Y^2 
     \right)^2 
   + \frac{1}{2 \lambda^0} 
     \left( 
      \del_0 Y^M - \lambda^1 \del_1 Y^M 
     \right)^2 
  \right. 
 } \nonumber \\ 
 && \qquad \quad 
 \displaystyle{ 
  \left.
  - \frac{\lambda^0}{2} (\del_1 Y^M)^2 
  \right]
     \, , 
 } \label{eq:action_now} 
\end{eqnarray} 
where $d^2 \sigma \equiv d \sigma^0 d \sigma^1$. 
 The symmetry of the action $S_0$ is 
given by the dimensional reduction of 
(\ref{eq:gauge_transformation}), 
except for the transformation law for $Y^2$, 
\begin{equation} 
 \delta_\epsilon Y^2 
  = \frac{1}{g} \widehat{\epsilon}^{\,2} 
    + \widehat{\epsilon}^{\,j} \del_j Y^2 
     \, , 
\end{equation} 
where $j = 0, 1$. 
 We can fix the symmetry corresponding 
to the parameter $\widehat{\epsilon}^{\,2}$ 
by taking $Y^2=0$. 
 Such a gauge fixing does not invoke any dynamical ghost field. 
 After a Gaussian integration over $\lambda^2$, 
we end up with the action in eq.~(\ref{barcelosstring}). 
 In this form, 
it is a familiar procedure to construct the BRS charge 
and examine if it is nilpotent or not and find that 
the BRS symmetry is anomalous 
unless the space-time dimension is $26$. 
 This fact indirectly implies that the critical dimension 
of the bosonic membrane is $27$. 
 From the membrane theory point of view, 
the anomaly can appear 
because the metrics (\ref{eq:metric_gamma}) and 
(\ref{eq:three_metric}) are singular in 
the background (\ref{eq:classical_bg_2}). 

 Trying to study the membrane theory more directly, 
let us recover the radius $R$ of the compactified circle 
and discretize it with a finite cutoff length $a$. 
This cutoff regularizes the divergences in the perturbation theory. 
 The membrane can then be regarded as a collection of 
Polyakov-like strings placed on this circle, 
each of which interacts with its neighbors. 
 For each Polyakov string to be consistent, 
we may be able to 
conclude that the critical dimension of bosonic membrane theory 
is $27$. 
 It is beyond the scope of this paper to investigate in detail 
whether this model gives a 
nonperturbative definition of the world-volume theory of membranes 
by taking an appropriate nontrivial continuum scaling limit. 

\section{Dimensional reduction of time-like direction} 
\label{sec:time_reduction} 
\subsection{Schild string theory} 
\label{sec:bosonic_sl_string} 

 Since the coordinates $\sigma^0$ and $\sigma^r$ are not 
on the equal footing in the action in eq.~(\ref{eq:action_main}), 
there exists a way of dimensional reduction inequivalent to the one 
in the previous section. 
 Namely we can wrap the membrane along the coordinate $\sigma^0$. 
 Let us describe the string theory we can get as the result of 
such a dimensional reduction. 

 In this section, 
we consider the world-volume with the Euclidean signature. 
 Since $\lambda^0$ is like an einbein field, 
this is achieved by replacing 
$\lambda^0$ and its antifield $\lambda^*_0$ as 
\begin{equation} 
 \lambda^0 \rightarrow -i \lambda^0 \, , \quad 
 \lambda^*_0 \rightarrow i \lambda^*_0 \, . 
\end{equation} 
 The Euclidean action 
is then obtained by multiplying the Minkowskian action by $-i$. 
 We will use the name $\sigma^0,~\lambda^0$ and $\lambda^*_0$ 
even after Euclideanization. 

 Let us wrap 
the $\sigma^0$-direction around 
a circularly compactified space-time direction $X^0$, 
\begin{equation} 
 X^0_{(0)} = \sigma^0 \, , \quad 
 X^I_{(0)} = 0 \quad (I = 1, \cdots, D-1) 
  \, , 
\end{equation} 
and consider the string model obtained 
by taking the radius of the circle $R\rightarrow 0$. 
 In this process, 
the coupling constant $g_2^2 = g^2 /(2\pi R)$ 
in the two-dimensional theory 
is kept fixed, 
which is redefined hereafter as new $g^2$. 
 The above configuration itself 
is not a classical solution of the original action 
$S_0$ in eq.~(\ref{eq:action_main}), but later 
we will turn on further backgrounds 
so that the final configuration satisfies 
the equations of motion. 

 Now, in terms of the fluctuation $Y^\mu$ defined by 
\begin{eqnarray} 
 X^0 &=& \sigma^0 + g Y^0 \, , 
  \nonumber \\ 
 X^I &=& g Y^I \quad (I = 1, \cdots, D-1) \, , 
\end{eqnarray} 
the action becomes 
\begin{eqnarray} 
 S_0 &=& 
  \displaystyle{ 
   \int d^2 \sigma 
   \left[ 
    \frac{1}{2 \lambda^0} 
    \left( 
     \frac{1}{g} - \lambda^0 \omega^r \del_r Y^0 
    \right)^2 
   \right. 
  } \nonumber \\ 
 && \qquad \qquad 
 \displaystyle{ 
  \left. 
   + \frac{\lambda^0}{2} 
     \left( 
      \omega^r \del_r Y^I 
     \right)^2 
   + \frac{\lambda^0}{4} g^2 
     \left( 
      \left\{ Y^\mu, Y^\nu \right\} 
     \right)^2 
  \right] \, , 
 } \label{eq:action_just} 
\end{eqnarray} 
where $\omega^r \equiv \lambda^r /\lambda^0$ and 
$\mu ,~\nu =0,~\cdots ,~D-1$. 
 The transformation properties 
are expressed in terms of $\omega^r$ as 
\begin{eqnarray} 
 \delta_\epsilon Y^0 
 &=& 
 \displaystyle{ 
    \frac{1}{g} \widehat{\epsilon}^{\,0} 
  + \widehat{\epsilon}^{\,s} \del_s Y^0 \, , 
 } \nonumber \\ 
 \delta_\epsilon Y^I 
 &=& 
 \displaystyle{ 
  \widehat{\epsilon}^{\,s} \del_s Y^I \, , 
 } \nonumber \\ 
 \delta_\epsilon \lambda^0 
 &=& 
 \displaystyle{ 
  - \lambda^0 \del_s \widehat{\epsilon}^{\,s} 
  - 2 (\lambda^0)^2 \omega^s \del_s \widehat{\epsilon}^{\,0} 
  + \widehat{\epsilon}^{\,s} \del_s \lambda^0 \, , 
 } \nonumber \\ 
 \delta_\epsilon \omega^r 
 &=& 
 \displaystyle{ 
     \del_s (\omega^r \widehat{\epsilon}^{\,s}) 
   + \lambda^0 J^{rs} \del_s \widehat{\epsilon}^{\,0} 
   - \omega^s \del_s \widehat{\epsilon}^{\,r} 
    \, , 
 } \label{eq:Schild_symmetry} 
\end{eqnarray} 
where 
\begin{equation} 
 J^{rs} = h h^{rs} + \omega^r \omega^s \, . 
\end{equation} 

 If one fixes the symmetry corresponding to 
$\widehat{\epsilon}^{\,0}$ by taking $Y^0=0$, the action becomes 
\beq 
 S_0= 
  \displaystyle{ 
   \int d^2 \sigma 
   \left[ 
    \frac{1}{2 \lambda^0g^2}
   + \frac{\lambda^0}{2} 
     \left( 
      \omega^r \del_r Y^I 
     \right)^2 
   + \frac{\lambda^0}{4} g^2 
     \left( 
      \left\{ Y^I, Y^J \right\} 
     \right)^2 
  \right] \, .
 } 
\eeq 
 Here $I,~J=1,\cdots ,D-1$. 
 Now, since $\omega^r$ are just auxiliary fields, 
we can integrate them out and we obtain 
\beq 
 S_0= 
  \displaystyle{ 
   \int d^2 \sigma 
   \left[ 
    \frac{1}{2 \lambda^0g^2} 
   + \frac{\lambda^0g^2}{4} 
     \left( 
      \left\{ Y^I, Y^J \right\} 
     \right)^2 
  \right] \, , 
 } 
  \label{Schild} 
\eeq 
which is nothing but the Schild action in a $D-1$ dimensional space-time\cite{Schild}. 
 Therefore we can obtain Schild string 
by dimensionally reducing our membrane action. 

 In the rest of this section, we would like to demonstrate the existence of 
the critical dimension starting from the action 
in eq.~(\ref{eq:action_just}). 
 Since this action is equivalent to the Schild string action,  
what we will show gives a way to get the critical dimension 
of the Schild string theory. 
 The purpose of studying Schild string theory is twofold. 
 For one thing, 
we want to check if the critical dimension of the bosonic membrane 
is $27$ or not, by trying another way of dimensional reduction. 
 For another thing, by studying Schild strings, 
we may be able to get some clue as 
to how we can treat the membrane action perturbatively. 
 Like the membrane action in eq.~(\ref{eq:action_main}), 
we should find out an appropriate classical background 
in the Schild action (\ref{Schild}) 
to start perturbative expansions. 
 Usually, a BRS covariant nondegenerate metric 
is available in such a background. 
 However, starting from the action (\ref{eq:action_just}), 
which is a simple modification of eq.~(\ref{Schild}), 
one can deduce the critical dimension. 
 Therefore we can expect that there exists 
a modification of eq.~(\ref{eq:action_main}) 
from which one can deduce the critical dimension. 

 For our purpose, 
it is convenient to fix  the gauge symmetry 
corresponding to $\widehat{\epsilon}^{\,0}$ 
by a gauge fixing condition $\del_r \omega^r = 0$. 
 Then, $\omega^r$ should take the form 
\begin{equation} 
 \omega^r 
  = g \epsilon^{rs} \del_s Y \, , 
   \label{eq:omega_Y} 
\end{equation} 
and the action becomes 
\begin{eqnarray} 
 S_0 
 &=& 
 \displaystyle{ 
  \int d^2 \sigma \, 
  \left[ 
   \frac{1}{2 \lambda^0} 
   \left( 
      \frac{1}{g} 
    + g \lambda^0 \left\{ Y, Y^0 \right\} 
   \right)^2 
  \right. 
 } \nonumber \\ 
 && \qquad \qquad 
 \displaystyle{ 
  \left. 
   + \frac{\lambda^0}{2} 
     \left( 
       g \left\{ Y, Y^I \right\} 
     \right)^2 
   + \lambda^0\,\frac{g^2}{4}\, 
     \left( \left\{ Y^\mu, Y^\nu \right\} \right)^2 
  \right] \, . 
 } \label{eq:Schild_type} 
\end{eqnarray} 
We will consider a classical background 
\begin{equation} 
 g Y_{(0)} = \sigma^2 \, , \quad 
 g Y^0_{(0)} = \sigma^1 \, .  
\label{Schild background}
\end{equation} 
 The kinetic terms for 
all the fields around this background are not pathological. 
 In this background, 
the induced metric $\partial_rY^I\partial_sY^I$ 
is singular. 
 It may appear that one can construct a metric using 
$Y$ and $Y^0$ but such a metric does not transform properly under 
the symmetry corresponding to $\widehat{\epsilon}^{\,0}$. 
 This action also looks like a Schild action but with 
two more coordinates $Y$, $Y^0$ compared to eq.~(\ref{Schild}). 
 Supersymmetrization of this action 
may be relevant to F-theory \cite{F_theory}. 
 We examine the perturbative quantum dynamics starting from 
this action, 
in particular focusing on 
the quantum-mechanical consistency of the gauge symmetries 
(\ref{eq:Schild_symmetry}). 

\subsection{Anomaly and critical dimension} 
\label{subsec:anomaly_bosonic_Schild} 
 After further fixing the symmetry by 
a condition $\lambda^0 = 1$, 
the dynamics around the background in eq.~(\ref{Schild background}) 
with respect to the fluctuation $A_r$ defined by 
\begin{eqnarray} 
 Y &=& 
 \displaystyle{ 
  \frac{1}{g} \sigma^2 - A_1 \, , 
 } \nonumber \\ 
 Y^0 &=& 
 \displaystyle{ 
  \frac{1}{g} \sigma^1 + A_2 \, , 
 } 
\end{eqnarray} 
is described by 
the action, 
\begin{equation} 
 S_{\rm matter} 
 = 
 \int d^2 \sigma \, 
 \left[ 
    \frac{1}{4} \left( F_{rs} \right)^2 
  + \frac{1}{2} \left( D_r Y^I \right)^2 
  + \frac{g^2}{4} 
    \left( 
     \{ Y^I, Y^J \} 
    \right)^2 
 \right] \, , 
  \label{eq:Yang_Mills} 
\end{equation} 
where 
\begin{eqnarray} 
 &\displaystyle{ 
  F_{rs} 
  \equiv 
  \del_r A_s - \del_s A_r 
    + g \left\{ A_r, A_s \right\} \, , 
  }& \nonumber \\ 
 &\displaystyle{ 
  D_r Y^I \equiv 
  \del_r Y^I + g \left\{ A_r, Y^I \right\} \, . 
 }& 
\end{eqnarray} 
 This action looks 
like the Yang-Mills action and it is invariant 
under the area-preserving diffeomorphism. 
 To fix the area-preserving diffeomorphism, 
we will take a covariant gauge fixing term 
mimicking the Yang-Mills case 
to respect the covariance of 
the free parts of $A_r$ in eq.~(\ref{eq:Yang_Mills}). 

 To analyze this system perturbatively,   
we provide the gauge fixing and ghost terms 
\`a la Batalin-Vilkoviski procedure. 
 The action satisfying the classical master equation 
can be obtained by either 
dimensionally reducing the action (\ref{eq:BV_membrane_action}) 
or starting 
from the dimensionally reduced action (\ref{eq:action_just}). 
 Both give the same result. 
 In order to impose the gauge fixing conditions described above, 
it is convenient to introduce a ghost field $C^{-1}$ and 
its antifield $C^*_{-1}$, 
an antighost field $\overline{C}_{-1}$ and 
its antifield $\overline{C}^{*\, -1}$, 
auxiliary fields $B,~\overline{\lambda}_0,~\overline{\omega}_r$ 
and their antifields 
$B^*,~\overline{\lambda}^{*\,0},~\overline{\omega}^{*\,r}$, 
and add the following term to the action: 
\begin{equation} 
 S^E_{-1} 
 = 
 \int d^2 \sigma \, 
 \left( 
  - i Y^* C^{-1} 
  + \overline{C}^{*\,-1} B 
  + \overline{C}^{*\,0} \overline{\lambda}_0 
  + \overline{C}^{*\,r} \overline{\omega}_r 
 \right) \, . 
\end{equation} 
 By taking the gauge fermion $\psi$, 
\begin{eqnarray} 
 \psi 
 &=& 
 \displaystyle{ 
   \int d^2 \sigma \, 
   \left[ 
    \overline{C}_{-1} 
    \left( 
       \del_1 Y - \frac{1}{g} \del_2 X^0 
     - \frac{\alpha}{2} B 
    \right) 
    + \overline{C}_0 \left( \lambda^0 - 1 \right) 
   \right. 
 } \nonumber \\ 
 && \qquad \qquad 
 \displaystyle{ 
   \left. 
    + 
    \overline{C}_r 
    \left( 
     \frac{1}{g} \omega^r - \epsilon^{rs} \del_s Y 
    \right) 
   \right] 
 } \nonumber \\ 
 &=& 
 \displaystyle{ 
  \int d^2 \sigma \, 
  \left[ 
   \overline{C}_{-1} 
   \left( 
    - \del_r A_r - \frac{\alpha}{2} B 
   \right) 
   + \overline{C}_0 \left( \lambda^0 - 1 \right) 
   + 
   \overline{C}_r 
   \left( 
    \frac{1}{g} \omega^r - \epsilon^{rs} \del_s Y 
   \right) 
  \right] \, ,
 } 
\end{eqnarray} 
the antifields are fixed as follows:
\begin{eqnarray} 
 && 
 \displaystyle{ 
  \overline{C}^{*\,-1} 
  = - \left( 
       \del_r A_r + \frac{\alpha}{2} B 
      \right) \, , 
 } \nonumber \\ 
 && 
 \displaystyle{ 
  Y^* 
  = - \del_1 \overline{C}_{-1} 
    + \epsilon^{rs} \del_s \overline{C}_r 
     \, , 
 } \nonumber \\ 
 && 
 \displaystyle{ 
  X^*_0 = 
  \frac{1}{g}\, \del_2 \overline{C}_{-1} \, , 
 } \nonumber \\ 
 && 
 \displaystyle{ 
  \omega^*_r = \overline{C}_r \, , 
 } \nonumber \\ 
 && 
 \displaystyle{ 
  \lambda^*_0 = \overline{C}_0 \, , 
 } \nonumber \\ 
 && 
 \displaystyle{ 
  \overline{C}^{*\,r} = 
   \frac{1}{g} \omega^r - \epsilon^{rs} \del_s Y \, , 
 } \nonumber \\ 
 && 
 \displaystyle{ 
  \overline{C}^{*\,0} = \lambda^0 - 1 \, . 
 } 
\end{eqnarray} 
 Hence the action $S^E_{-1}$ becomes 
\begin{eqnarray} 
 S^E_{-1\,\psi} 
 &\equiv& 
 \displaystyle{ 
  \left. 
   S^E_{-1} 
  \right|_{\phi^*_i = \del \psi /\del \phi^i} 
 } \nonumber \\ 
 &=& 
 \displaystyle{ 
  \int d^2 \sigma 
  \left[ 
     i \del_1 \overline{C}_{-1} C^{-1} 
   + i \left( 
        \epsilon^{rs} \del_r \overline{C}_s 
       \right) C^{-1} 
  \right. 
 } \nonumber \\ 
 && \qquad \quad 
 \displaystyle{ 
  \left. 
   + 
   \left( 
    - \del_r A_r - \frac{\alpha}{2} B 
   \right) B 
  \right. 
 } \nonumber \\ 
 && \qquad \quad 
 \displaystyle{ 
  \left. 
   + \overline{\lambda}_0 
     \left( 
      \lambda^0 - 1 
     \right) 
   + \overline{\omega}_r 
     \left( 
      \frac{1}{g} \omega^r - \epsilon^{rs} \del_s Y 
     \right) 
  \right] \, . 
 } \label{eq:S_psi} 
\end{eqnarray} 
 Integrating over the auxiliary fields 
$B,~\overline{\lambda}_0,~\overline{\omega}_r$, 
we get the gauge fixing conditions 
$\lambda^0 = 1$,  
$\omega^r = g \epsilon^{rs} \del_s Y$, 
and a covariant gauge fixing term 
\begin{equation} 
 S_{\rm GF} 
 = \int d^2 \sigma\, \frac{1}{2 \alpha} 
   \left( \del_r A_r \right)^2 \, . 
\end{equation} 
 Finally we get the action 
\beq 
 S_{\rm matter} + S_{\rm GF} + S_{\rm ghost}, 
  \label{eq:total_action} 
\eeq 
where 
\begin{eqnarray} 
 S_{\rm ghost} &=& 
\displaystyle{
  \int d^2 \sigma 
  \left[ 
     i \del_1 \overline{C}_{-1} C^{-1} 
   + i \left( 
        \epsilon^{rs} \del_r \overline{C}_s 
       \right) C^{-1}
  \right.
 } \nonumber \\
 && \qquad
 \displaystyle{ 
  \left.
   - i \del_2 \overline{C}_{-1} C^0 
   - i \del_2 \overline{C}_{-1} C^1 
  \right. 
 } \nonumber \\ 
 && \qquad 
 \displaystyle{ 
   + 2 i \overline{C}_0 \del_1 C^0 
   + i \overline{C}_0 \del_r C^r 
 } \nonumber \\ 
 && \qquad 
 \displaystyle{ 
  - i \overline{C}_r \del_r C^0 
  + i \overline{C}_r \del_1 C^r 
  - i \overline{C}_1 \del_s C^s 
 } \nonumber \\ 
 && \qquad 
 \displaystyle{ 
  - i g\, \del_2 \overline{C}_{-1} (\del_r A_2) C^r 
  + 2 i g\, \overline{C}_0 
    \left\{ A_1, C^0 \right\} 
 } \nonumber \\ 
 && \qquad 
 \displaystyle{ 
  - i g\, \overline{C}_r \left\{ A_r, C^0 \right\} 
  + i g\, \overline{C}_r \epsilon^{rs} (\del_s A_t) 
    \del_t C^0 
 } \nonumber \\ 
 && \qquad 
 \displaystyle{ 
  + i g\, \overline{C}_r \left\{ A_1, C^r \right\} 
  - i g\, (\del_s \overline{C}_r) \epsilon^{rt} (\del_t A_1) C^s 
 } \nonumber \\ 
 && \qquad 
 \displaystyle{ 
  + i g^2\,\overline{C}_r \epsilon^{rs} (\del_s A_t) 
    \left\{ A_t, C^0 \right\} 
  + i g^2\,\overline{C}_r \epsilon^{rs} \del_s Y^I 
    \left\{ 
     Y^I, C^0 
    \right\} 
 } \nonumber \\ 
 && \qquad 
 \displaystyle{ 
  \left. 
   + \frac{3}{4}\, g^2\, 
     \epsilon^{tu} \overline{C}_t \overline{C}_u 
     \left\{ C^0, C^0 \right\} 
  \right] \, . 
 } \label{eq:ghost_for_Schild} 
\end{eqnarray} 

 Now that 
we have the gauge-fixed action which is BRS invariant at least 
classically, 
the most honest way to examine if the BRS symmetry is 
anomalous or not is to check 
if the BRS charge is nilpotent quantum mechanically. 
 Here we take a by-pass which is 
usually taken in this kind of situation. 
 We will define the quantities $j_0$ and $j_r$ 
which we will call currents as the variations of 
$\overline{C}_0 $ and $\overline{C}_r$ under the 
BRS transformation as
\bea 
 & 
 \displaystyle{ 
  j_0 \equiv s \overline{C}_0 \, , 
 }& \nonumber \\ 
 & 
 \displaystyle{ 
  j_r \equiv s \overline{C}_r \quad (r = 1, 2) \, . 
 }& 
  \label{eq:current_def} 
\eea 
 Essentially $j_0$ and $j_r$ are 
the BRS invariant version of the constraints $\phi_0$ and $\phi_r$ 
in the original action. 
 Therefore, by checking if there is any Schwinger terms 
in the algebra satisfied by $j_0$ and $j_r$, 
we can see if the BRS symmetry is anomalous or not. 

 There is another way to look at these currents. 
 Since $\overline{C}_0$ and $\overline{C}_r$ are 
the antifields of $\lambda^0$ and $\omega^r$, 
$j_0$ and $j_r$ can be written as 
\bea 
&& 
 \displaystyle{ 
 \left. 
  j_0 = \frac{\delta S}{\delta\lambda^0} 
 \right|_{\phi^*_i = \del \psi /\del \phi^i} \, , 
 } \nonumber \\ 
&&
 \displaystyle{ 
 \left. 
  j_r = \frac{\delta S}{\delta\omega^r} 
 \right|_{\phi^*_i = \del \psi /\del \phi^i} \, , 
 } 
\eea 
where $S$ is the action before the gauge fixing. 
 Therefore the currents can be considered as representing 
the response of the action 
to variations of the gauge fixing conditions for $\lambda^0$ 
and $\omega^r$. 
 These currents play a role similar to the one played by 
the energy-momentum tensor in the Polyakov string theory. 
 If the reparametrization symmetry is not anomalous, 
the correlation functions of these currents should vanish 
up to local terms proportional to derivatives of delta functions 
in the position space.  
 
 The explicit form of these currents is found to be 
$j_a = 
   \left. j_a \right|_{\rm matter} 
 + \left. j_a \right|_{\rm ghost}$, 
where 
the matter contributions $\left. j_a \right|_{\rm matter}$ are  
\begin{eqnarray} 
 \left. j_0 \right|_{\rm matter} 
 &=& 
 \displaystyle{ 
  - \frac{1}{2g^2} 
  + \frac{g^2}{2} 
    \left( \{ Y, Y^0 \} \right)^2 
  + \frac{g^2}{2} \left( \{ Y, Y^I \} \right)^2 
  + \frac{g^2}{4} 
    \left( \left\{ Y^\mu, Y^\nu \right\} \right)^2 
 } \nonumber \\ 
 &=& 
 \displaystyle{ 
  \frac{1}{g} \epsilon^{rs} \del_r A_s 
 } \nonumber \\ 
 && \ 
 \displaystyle{ 
  + \frac{1}{4} \left( 
                  \del_r A_s - \del_s A_r 
                \right)^2 
  + \frac{1}{2} \epsilon^{rs} 
    \left\{ A_r, A_s \right\} 
  + \frac{1}{2} (\del_r Y^I)^2 
  + {\cal O}(g) \, , 
 } \nonumber \\ 
 \left. j_r \right|_{\rm matter} 
 &=& 
 \displaystyle{ 
  - \frac{1}{g} \del_r Y^0 
  - g \del_r Y^\mu 
      \left\{ Y, Y_\mu \right\} 
 } \nonumber \\ 
 &=& 
 \displaystyle{ 
    \delta_{r1}\,\frac{1}{g} \epsilon^{st} \del_s A_t 
 } \nonumber \\ 
 && \ 
 \displaystyle{ 
  + (\del_r A_2) \epsilon^{st} \del_s A_t 
  + \delta_{r1}\, 
    \frac{1}{2} \epsilon^{st} \left\{ A_s, A_t \right\} 
  + \del_r Y^I \del_1 Y^I 
  + {\cal O}(g) \, , 
 } 
  \label{eq:current_matter} 
\end{eqnarray} 
while the ghost contributions 
$\left. j_a \right|_{\rm ghost}$ are 
\begin{eqnarray} 
 \left. j_0 \right|_{\rm ghost} 
 &=& 
 \displaystyle{ 
    4 i \overline{C}_0 \del_1 C^0 
  + 2 i \overline{C}_0 \del_s C^s 
  + i \del_s \overline{C}_0 C^s 
  - i \overline{C}_r \del_r C^0 
 } \nonumber \\ 
 && \ 
 \displaystyle{ 
  + 4 i g\, \overline{C}_0 
    \left\{ A_1, C^0 \right\} 
  - i g\, \overline{C}_r \left\{ A_r, C^0 \right\} 
  + i g\, \overline{C}_r 
    \epsilon^{rs} (\del_s A_t) \del_t C^0 
 } \nonumber \\ 
 && \ 
 \displaystyle{ 
  + i g^2\, \overline{C}_r \epsilon^{rs} (\del_s A_t) 
    \left\{ A_t, C^0 \right\} 
  + i g^2\, \overline{C}_r \epsilon^{rs} (\del_s Y^I) 
    \left\{ Y^I, C^0 \right\} 
 } \nonumber \\ 
 && \ 
 \displaystyle{ 
  + \frac{3}{4}\, g^2\, 
    \epsilon^{tu} \overline{C}_t \overline{C}_u 
    \left\{ C^0, C^0 \right\} \, , 
 } \nonumber \\ 
 \left. j_r \right|_{\rm ghost} 
 &=& 
 \displaystyle{ 
    2 i \overline{C}_0 \del_r C^0 
  - i \overline{C}_r \del_1 C^0 
  - i \overline{C}_1 \del_r C^0 
  + i \overline{C}_s \del_r C^s 
  + i (\del_s \overline{C}_r) C^s 
 } \nonumber \\ 
 && \ 
 \displaystyle{ 
  - i g\, \overline{C}_r 
    \left\{ A_1 , C^0 \right\} 
  + i g\, \overline{C}_s \epsilon^{st} 
   (\del_t A_1) \del_r C^0 \, . 
 } 
\end{eqnarray} 
 We will calculate the two-point functions of these currents to 
see if there is any anomaly. 
 The local terms in such two-point functions proportional to 
derivatives of the delta function in position space 
are irrelevant, 
because such terms can be cancelled 
by adding appropriate local counterterms. 
 If the nonlocal terms remain even after 
the addition of any possible counterterms to 
the bare action and the currents,  
the symmetry is anomalous. 

 It is easy to check that 
the tree level contributions to 
the current-current correlation functions are local. 
 Thus, we are interested in the one-loop correction. 
 However, 
as seen from the bilinear part of 
(\ref{eq:S_psi}) and (\ref{eq:ghost_for_Schild}), 
the various ghosts and antighosts mix in a complicated way, 
making all the calculations rather cumbersome. 
 Thus, we pose here to make the ghost sector as simple as possible. 
 We note that 
the ghost ${C}^{-1}$ appears 
only in the first line of eq.~(\ref{eq:S_psi}) 
and does not appear in the currents 
to the relevant order of $g$. 
 Writing 
\begin{equation} 
 \overline{C}_1^\prime \equiv \overline{C}_1\, , 
  \quad 
 \overline{C}_2^\prime 
  \equiv \overline{C}_2 + \overline{C}_{-1} \, , 
\end{equation} 
integration over ${C}^{-1}$ gives 
\begin{equation} 
 \epsilon^{rs} \del_r \overline{C}_s^\prime = 0 \, . 
\end{equation} 
 If we consider only world-sheets with trivial topology, 
this condition is solved by introducing a new variable $b$ as 
\begin{equation} 
 \overline{C}_r^\prime = i \del_r b \, . 
\end{equation} 
 Hence, the original variables $\overline{C}_r$ 
are expressed in terms of $b$ and $\overline{C}_{-1}$ as 
\begin{eqnarray} 
 \overline{C}_r &=& 
  i \del_r b - \delta_{r 2}\, \overline{C}_{-1} \, . 
\end{eqnarray} 
 The ghost sector now consists of 
three pairs of ghost and antighost. 
 Further change of the variables from 
$C^r$ to $E^r$ 
\begin{equation} 
 C^r = E^r -2 \delta^{r 1}\, C^0 
       + 2 g (\epsilon^{rs} \del_s A_1) C^0 \, , 
\end{equation} 
simplifies the structure of interactions. 
 Now the free propagators of the ghost variables 
are found as 
\begin{eqnarray} 
 && 
 \displaystyle{ 
  \left( 
   \begin{array}{ccc} 
    \left< 
     C^0(\sigma) b(\sigma^\prime) 
    \right> & 
    \left< 
     C^0(\sigma) \overline{C}_0(\sigma^\prime) 
    \right> & 
    \left< 
     C^0(\sigma) \overline{C}_{-1}(\sigma^\prime) 
    \right> 
     \\ 
    \left< 
     E^1(\sigma) b(\sigma^\prime) 
    \right> & 
    \left< 
     E^1(\sigma) \overline{C}_0(\sigma^\prime) 
    \right> & 
    \left< 
     E^1(\sigma) \overline{C}_{-1}(\sigma^\prime) 
    \right> 
     \\ 
    \left< 
     E^2(\sigma) b(\sigma^\prime) 
    \right> & 
    \left< 
     E^2(\sigma) \overline{C}_0(\sigma^\prime) 
    \right> & 
    \left< 
     E^2(\sigma) \overline{C}_{-1}(\sigma^\prime) 
    \right> 
   \end{array} 
  \right) 
 } \nonumber \\ 
 && \qquad 
 \displaystyle{ 
  = 
  \int \frac{d^2 q}{(2\pi)^2} 
  \, e^{i q \cdot (\sigma - \sigma^\prime)} \, 
  \frac{1}{q^2} 
  \left( 
   \begin{array}{ccc} 
      1 &    0 &     0 \\ 
      0 & - q_1 & - q_2 \\ 
      0 & - q_2 &   q_1 
   \end{array} 
  \right) \, . 
 } 
\end{eqnarray} 
\begin{figure}[htb] 
\begin{center} 
\begin{picture}(300, 100)(0, 0) 
 \SetWidth{1.0} 
 \CArc(150, 50)(40, 0, 180) 
 \CArc(150, 50)(40, 180, 360) 
 \Line(50,50)(110,50) 
 \Line(190,50)(250,50) 
 \Vertex(110, 50){4} 
 \Vertex(190, 50){4} 
 \Vertex(50,50){4} 
 \Vertex(250,50){4} 
 \Text(40, 50)[r]{{\Large $j_a$}} 
 \Text(260, 50)[l]{{\Large $j_b$}} 
\end{picture} 
\\ 
{\rm (a) Vacuum polarization contribution} 
\end{center} 
\begin{center} 
\begin{picture}(300, 100)(0, 0) 
 \SetWidth{1.0} 
 \CArc(70, 50)(40, 0, 180) 
 \CArc(70, 50)(40, 180, 360) 
 \Line(-10, 50)(30, 50) 
 \Vertex(30, 50){4} 
 \Vertex(-10, 50){4} 
 \Vertex(110, 50){4} 
 \Text(-20, 50)[r]{{\Large $j_a$}} 
 \Text(120, 50)[l]{{\Large $j_b$}} 
%
%
 \CArc(220, 50)(40, 0, 180) 
 \CArc(220, 50)(40, 180, 360) 
 \Line(260, 50)(300, 50) 
 \Vertex(180, 50){4} 
 \Vertex(260, 50){4} 
 \Vertex(300, 50){4} 
 \Text(170, 50)[r]{{\Large $j_a$}} 
 \Text(310, 50)[l]{{\Large $j_b$}} 
\end{picture} 
\\ 
{\rm (b) One propagator contribution} 
\end{center} 
\begin{center} 
\begin{picture}(300, 100)(0, 0) 
 \SetWidth{1.0} 
 \CArc(150, 8)(80, 30, 150) 
 \CArc(150, 90)(80, 210, 330) 
 \Vertex(80, 50){4} 
 \Vertex(220, 50){4} 
 \Text(70, 50)[r]{{\Large $j_a$}} 
 \Text(230, 50)[l]{{\Large $j_b$}} 
\end{picture} 
\\ 
{\rm (c) Two-particle irreducible contribution} 
\end{center} 
\caption{One-loop diagrams in current-current correlation} 
\vspace{1.0cm} 
\label{fig:diagram} 
\end{figure} 

 In the present example, 
the one-loop quantum correction 
to the two-point current correlation functions 
consists of three types of Feynman diagrams 
shown in Fig.~\ref{fig:diagram}. 
 We demonstrate 
the existence of an anomaly except for $D=27$ 
at the one-loop level 
by evaluating 
the nonlocal part of the simplest two-point function 
\begin{equation} 
 \Pi_{22}(p) 
 \equiv 
 \left. 
  \int d^2 \sigma\, 
  e^{-i p \cdot \sigma} 
  \left< j_2(\sigma) j_2(0) \right> 
 \right|_{\rm nonlocal} \, . 
  \label{eq:def_Pi_22} 
\end{equation} 
 Since $j_2$ involves no terms linear in fields, 
the two-particle irreducible diagram 
shown in Fig.~\ref{fig:diagram}(c) gives the lowest order 
contribution. 
 In perturbation theory, the effect of adding counterterms 
is a higher order effect. 
 Thus, if this lowest order contribution is nonzero, 
the theory is anomalous. 
 From eq.~(\ref{eq:current_matter}), 
we find the nonlocal contributions to 
$\Pi_{22}(p)$ 
from the scalar $Y^I$ and the gauge field $A_r$ as 
\footnote{ 
 We use the Feynman gauge, $\alpha = 1$ in the calculation. 
} 
\begin{eqnarray} 
 \displaystyle{ 
  \left. 
   \Pi_{22}(p) 
  \right|_{Y^I} 
 } 
 &=& 
 \displaystyle{ 
  \frac{1}{4\pi} 
  \frac{D-1}{3} 
  \frac{(p_1)^2 (p_2)^2}{p^2} \, , 
 } \nonumber \\ 
 \displaystyle{ 
  \left. 
   \Pi_{22}(p) 
  \right|_{A_r} 
 } 
 &=& 
 \displaystyle{ 
  \frac{1}{4\pi} 
  \left[ 
     \frac{p^2}{8} 
      {\rm ln}\left( \frac{p^2}{\mu^2} \right) 
   + \frac{1}{6} \frac{(p_1)^2 (p_2)^2}{p^2} 
  \right] \, , 
 } \label{eq:22_A_Y} 
\end{eqnarray} 
where $\mu$ is an arbitrary scale to be chosen as the 
renormalization point. 
 The ghost contribution to $\Pi_{22}(p)$ becomes 
\begin{equation} 
 \displaystyle{ 
  \left. 
   \Pi_{22}(p) 
  \right|_{\rm G} 
 } 
 = 
 \displaystyle{ 
  \frac{1}{4\pi} 
  \left[
   - \frac{p^2}{8} 
     {\rm ln}\left( \frac{p^2}{\mu^2} \right) 
   - \frac{53}{6} \frac{(p_1)^2 (p_2)^2}{p^2} 
  \right] \, . 
 } \label{eq:22_ghost} 
\end{equation} 
 The logarithmic term in this ghost contribution 
cancels that in the gauge boson contribution. 
 Now we find that 
the sum of (\ref{eq:22_A_Y}) and (\ref{eq:22_ghost}) becomes 
\begin{equation} 
 \displaystyle{ 
  \Pi_{22}(p) 
  = 
  \frac{1}{4\pi} \frac{D - 27}{3} 
   \frac{(p_1)^2 (p_2)^2}{p^2} \, . 
 } 
\end{equation} 
 The nonlocal terms arising from Fig.~\ref{fig:diagram}(c) 
cannot be removed by an addition of any local counterterms 
to the action and the currents, 
showing that the gauge symmetry 
is anomalous except for $D = 27$ in this order of approximation. 
 We need to check that the other two-point functions 
vanish if and only if $D$ = 27. 
 This fact is demonstrated in Appendix B. 
 The result we get implies that the critical dimension of 
the bosonic membrane is $27$ 
and the critical dimension of the Schild string is $26$, 
at least to this order of approximation. 

\section{Discussion} 
\label{sec:discussion} 

 In the first part of this paper, 
we have explored a perturbative regime 
of the membrane world-volume theory 
from Lagrangian and path integral point of view, 
in order to examine the existence of a potential anomaly 
in gauge symmetries. 
 The perturbation theory is defined 
only around an appropriate membrane background 
which guarantees the existence of non-pathological propagators 
for all dynamical degrees of freedom. 
 On the other hand, 
the gauge anomaly can appear only in the background 
where BRS covariant metrics are degenerate  
so that neither BRS invariant measures nor local counterterms  
cancelling anomalies are allowed. As far as we checked, 
 all the classical solutions with nondegenerate metrics 
do not induce perturbative membranes with an anomaly. 
 Even if we can check that some of the metrics 
are degenerate, there is always a possibility that 
there exists another BRS covariant metric we do not know 
which becomes nondegenerate. 
 Therefore a careful study is necessary to argue 
that the theory is anomalous around a background. 

 The analysis here does not 
contradict with the results given 
in Ref.~\cite{Marquard_Scholl_1,Marquard_Scholl_2}, 
which is based on 
the operator description. 
 The operator methods do not refer to any backgrounds. 
 One of the motivation 
to develop the perturbative method in this paper  
is to avoid being annoyed about the 
ambiguity in the operator ordering and regularization procedure. 
 Unfortunately we do not know any classical solutions 
of the membrane world-volume action examined here, 
with singular metrics which make the reparametrization symmetry 
anomalous. 
 
 In the second part, 
we examined two string models obtained by 
dimensional reduction of membranes. 
 Each of these string models 
accommodates an anomaly and critical dimension. 
 We may say that 
this aspect indicates that the original membrane theory also has 
the critical dimension, 
although it is of course much better to be derived 
in a three-dimensional theory. 
 Another possibility is that 
the critical dimension appears as 
the dimension where 
the regularized membrane model proposed 
in Sec.~\ref{sec:Polyakov_type} 
possesses a nontrivial continuum limit. 
 This limiting procedure may be related rather to 
the nonperturbative aspects of the membrane world-volume theory, 
which cannot be pursued 
by the perturbative analysis developed here. 

 In particular, 
we have found that the string theory with the Schild action 
possess the critical dimension $26$.  
 We expect that we can get the usual bosonic string theory 
also from the Schild action. 
 In order to confirm this, 
we should check if this string theory 
yields the same space-time equations of motion. 
$D-1$ $=$ $26$ is a part of the space-time equations of motion, 
and it is an intriguing problem to generalize this to 
get the whole equations of motion in the low energy approximation. 
 The Schild action is interesting 
because it enables one to take the tensionless limit.  
 Thus, by studying this action, 
we may be able to reveal the huge gauge symmetries 
of string theory. 

 Also the experience of dealing with the Schild action 
gives some hint as to how to deal with the membrane theory. 
 The reason why we were able to deduce the critical dimension 
for the Schild string is that we started from an action 
with some auxiliary fields and extra gauge symmetries. 
 These features originate from the fact that we started from 
a higher dimensional theory. 
 Therefore, an obvious strategy which should be tried 
to deal with the membrane theory is to consider 
an action obtained from reduction of higher dimensional branes.  
 Dimensional reduction of one spatial or time-like 
direction of a 3-brane is the simplest possibility. 
 Another possibility is to use the Born-Infeld action. 
 We hope to come back to this problem in the future.

 The matrix regularization of Schild string model 
is also interesting, 
recalling the role played by the Schild action in \cite{IKKT}. 
 The matrices can be considered as regularizing 
the area-preserving diffeomorphism. 
 Obviously, 
the matrix regularization is possible only when 
the action obtained after fixing the other gauge redundancy 
is written in terms of the Poisson bracket. 
 The matrix model should have the associated ghost sector 
so as to reproduce the critical dimension $D-1 = 26$ 
in the naive continuum limit. 

\section*{Acknowledgements} 

 We would like to thank K. Okuyama for discussions. 
 The work of N. I. is partially supported by 
Grant-in-Aid for Scientific Research (C) No. 13640308.

\section*{Appendix} 
\label{app:critical} 
 In this appendix, we collect the results 
for the one-loop correction 
to the current-current correlation functions, 
with the definition similar to (\ref{eq:def_Pi_22}) 
\begin{equation} 
 \Pi_{ab}(p) 
 \equiv 
 \left. 
  \int d^2 \sigma\, 
  e^{-i p \cdot \sigma} 
  \left< j_a(\sigma) j_b(0) \right> 
 \right|_{\rm nonlocal} \, , 
\end{equation} 
in the Schild-like string model. 
 Eq.~(\ref{eq:current_matter}) shows that 
$j_0(\sigma)$ and $j_1(\sigma)$ 
have the common terms linear in fields. 
 Thus, $j_- \equiv j_0 - j_1$ starts 
from the terms bilinear in fields, 
and we consider the two-point functions of 
$j_0$, $j_2$ and $j_-$. 

 For $\Pi_{--}(p)$ 
\begin{eqnarray} 
 \displaystyle{ 
  \left. 
   \Pi_{--}(p) 
  \right|_{Y^I} 
 } 
 &=& 
 \displaystyle{ 
  \frac{1}{4\pi} 
  \left( -\frac{D-1}{3} \right) 
  \frac{(p_1)^2 (p_2)^2}{p^2} \, , 
 } \nonumber \\ 
 \displaystyle{ 
  \left. 
   \Pi_{--}(p) 
  \right|_{A_r} 
 } 
 &=& 
 \displaystyle{ 
  \frac{1}{4\pi} 
  \left[ 
     \frac{p^2}{8} 
      {\rm ln}\left( \frac{p^2}{\mu^2} \right) 
   - \frac{1}{6} \frac{(p_1)^2 (p_2)^2}{p^2} 
  \right] \, , 
 } \nonumber \\ 
 \displaystyle{ 
  \left. 
   \Pi_{--}(p) 
  \right|_{\rm G} 
 } 
 &=& 
 \displaystyle{ 
  \frac{1}{4\pi} 
  \left[ 
   - \frac{p^2}{8} 
      {\rm ln}\left( \frac{p^2}{\mu^2} \right) 
   + \frac{53}{6} \frac{(p_1)^2 (p_2)^2}{p^2} 
  \right] \, , 
 } 
\end{eqnarray} 
which become in total 
\begin{equation} 
 \displaystyle{ 
  \Pi_{--}(p) 
 } 
 = 
 \displaystyle{ 
  \frac{1}{4\pi} 
  \frac{27 - D}{3} 
  \frac{(p_1)^2 (p_2)^2}{p^2} \, . 
 } 
\end{equation} 

 The vacuum polarization contribution 
$\left. \Pi_{00}(p) \right|^{(a)}$ to $\Pi_{00}(p)$ 
is found as 
\begin{eqnarray} 
 \displaystyle{ 
  \left. 
   \Pi_{00}(p) 
  \right|_{Y^I}^{(a)} 
 } 
 &=& 
 \displaystyle{ 
  \frac{1}{4\pi} \times 
  2 (D-1) \times 
  \frac{p^2}{8} 
  {\rm ln}\left( \frac{p^2}{\mu^2} \right) \, , 
 } \nonumber \\ 
 \displaystyle{ 
  \left. 
   \Pi_{00}(p) 
  \right|_{A_r}^{(a)} 
 } 
 &=& 
 \displaystyle{ 
  \frac{1}{4\pi} 
  (-7) 
  \frac{p^2}{8} 
  {\rm ln}\left( \frac{p^2}{\mu^2} \right)  \, , 
 } \nonumber \\ 
 \displaystyle{ 
  \left. 
   \Pi_{00}(p) 
  \right|_{\rm G}^{(a)} 
 } 
 &=& 
 \displaystyle{ 
  \frac{1}{4\pi} 
  (-5) 
  \frac{p^2}{8} {\rm ln}\left( \frac{p^2}{\mu^2} \right) 
   \, , 
 } 
\end{eqnarray} 
and thus, 
\begin{equation} 
 \displaystyle{ 
  \left. 
   \Pi_{00}(p) 
  \right|^{(a)} 
  = 
  \frac{1}{4\pi} 
  \left( - 12 + 2\left( D-1 \right) \right) 
  \frac{p^2}{8} {\rm ln}\left( \frac{p^2}{\mu^2} \right) 
   \, . 
    \label{eq:00_a} 
 } 
\end{equation} 
 The one propagator contribution 
$\left. \Pi_{00}(p) \right|^{(b)}$ to $\Pi_{00}(p)$ 
is found as 
\begin{eqnarray} 
 \displaystyle{ 
  \left. 
   \Pi_{00}(p) 
  \right|_{Y^I}^{(b)} 
 } 
 &=& 
 \displaystyle{ 
  \frac{1}{4\pi} 
  \left( -4 (D-1) \right) 
  \frac{p^2}{8} 
  {\rm ln}\left( \frac{p^2}{\mu^2} \right) 
   \, , 
 } \nonumber \\ 
 \displaystyle{ 
  \left. 
   \Pi_{00}(p) 
  \right|_{A_r}^{(b)} 
 } 
 &=& 
 \displaystyle{ 
  \frac{1}{4\pi} \times 16 \times 
  \frac{p^2}{8} {\rm ln} \left( \frac{p^2}{\mu^2} \right) 
    \, , 
 } \nonumber \\ 
 \displaystyle{ 
  \left. 
   \Pi_{00}(p) 
  \right|_{\rm G}^{(b)} 
 } 
 &=& 
 \displaystyle{ 
  \frac{1}{4\pi} 
  \times 12 \times 
  \frac{p^2}{8} {\rm ln}\left( \frac{p^2}{\mu^2} \right) 
   \, , 
 } 
\end{eqnarray} 
the sum of which becomes 
\begin{equation} 
 \displaystyle{ 
  \left. 
   \Pi_{00}(p) 
  \right|^{(b)} 
  = 
  \frac{1}{4\pi} 
  \left( 28 - 4\left( D-1 \right) \right) 
  \frac{p^2}{8} {\rm ln}\left( \frac{p^2}{\mu^2} \right) 
   \, . 
    \label{eq:00_b} 
 } 
\end{equation} 
 The two-particle irreducible contribution 
$\left. \Pi_{00}(p) \right|^{(c)}$ to $\Pi_{00}(p)$ 
is found as 
\begin{eqnarray} 
 \displaystyle{ 
  \left. 
   \Pi_{00}(p) 
  \right|_{Y^I}^{(c)} 
 } 
 &=& 
 \displaystyle{ 
  \frac{1}{4\pi} 
  \times 2 (D-1) \times  
  \frac{p^2}{8} 
  {\rm ln}\left( \frac{p^2}{\mu^2} \right) 
   \, , 
 } \nonumber \\ 
 \displaystyle{ 
  \left. 
   \Pi_{00}(p) 
  \right|_{A_r}^{(c)} 
 } 
 &=& 
 \displaystyle{ 
  \frac{1}{4\pi} (-8) 
  \frac{p^2}{8} {\rm ln} \left( \frac{p^2}{\mu^2} \right) 
    \, , 
 } \nonumber \\ 
 \displaystyle{ 
  \left. 
   \Pi_{00}(p) 
  \right|_{\rm G}^{(c)} 
 } 
 &=& 
 \displaystyle{ 
  \frac{1}{4\pi} 
  (-8) 
  \frac{p^2}{8} {\rm ln}\left( \frac{p^2}{\mu^2} \right) 
   \, , 
 } 
\end{eqnarray} 
and thus, 
\begin{equation} 
 \displaystyle{ 
  \left. 
   \Pi_{00}(p) 
  \right|^{(c)} 
  = 
  \frac{1}{4\pi} 
  \left( -16 + 2\left( D-1 \right) \right) 
  \frac{p^2}{8} {\rm ln}\left( \frac{p^2}{\mu^2} \right) 
   \, . 
    \label{eq:00_c} 
 } 
\end{equation} 
 The total contribution to $\Pi_{00}(p)$, 
the sum of 
(\ref{eq:00_a}), (\ref{eq:00_b}) and (\ref{eq:00_c}), 
vanishes; 
\begin{equation} 
 \Pi_{00}(p) = 0 \, . 
\end{equation} 

 There are potential contributions 
from Fig.~\ref{fig:diagram}(b) and (c) 
for $\Pi_{02}(p)$. 
 The one propagator contribution 
$\left. \Pi_{02}(p) \right|^{(b)}$ is found as 
\begin{eqnarray} 
 \displaystyle{ 
  \left. 
   \Pi_{02}(p) 
  \right|_{Y^I}^{(b)} 
 } 
 &=& 
 0 \, , 
 \nonumber \\ 
 \displaystyle{ 
  \left. 
   \Pi_{02}(p) 
  \right|_{A_r}^{(b)} 
 } 
 &=& 
 \displaystyle{ 
  \frac{1}{4\pi} \times 
  2 p_1 p_2 \times 
  \frac{1}{8} {\rm ln} \left( \frac{p^2}{\mu^2} \right) 
    \, , 
 } \nonumber \\ 
 \displaystyle{ 
  \left. 
   \Pi_{02}(p) 
  \right|_{\rm G}^{(b)} 
 } 
 &=& 
 \displaystyle{ 
  \frac{1}{4\pi} \times  
  6 p_1 p_2 \times 
  \frac{1}{8} {\rm ln}\left( \frac{p^2}{\mu^2} \right) 
   \, , 
 } 
\end{eqnarray} 
the sum of which gives 
\begin{equation} 
 \displaystyle{ 
  \left. 
   \Pi_{02}(p) 
  \right|^{(b)} 
  = 
  \frac{1}{4\pi} \times  
  8 p_1 p_2 \times 
  \frac{1}{8} {\rm ln}\left( \frac{p^2}{\mu^2} \right) 
   \, . 
 } \label{eq:02_b} 
\end{equation} 
 The two-particle irreducible contribution 
$\left. \Pi_{02}(p) \right|^{(c)}$ 
is found as 
\begin{eqnarray} 
 \displaystyle{ 
  \left. 
   \Pi_{02}(p) 
  \right|_{Y^I}^{(c)} 
 } 
 &=& 
 0 \, , 
  \nonumber \\ 
 \displaystyle{ 
  \left. 
   \Pi_{02}(p) 
  \right|_{A_r}^{(c)} 
 } 
 &=& 
 0 \, , 
  \nonumber \\ 
 \displaystyle{ 
  \left. 
   \Pi_{02}(p) 
  \right|_{\rm G}^{(c)} 
 } 
 &=& 
 \displaystyle{ 
  \frac{1}{4\pi} \, 
  (-8 p_1 p_2) \, 
  \frac{1}{8} {\rm ln}\left( \frac{p^2}{\mu^2} \right) 
   \, , 
 } 
\end{eqnarray} 
the sum of which becomes 
\begin{equation} 
 \displaystyle{ 
  \left. 
   \Pi_{02}(p) 
  \right|^{(c)} 
  = 
  \frac{1}{4\pi} \, 
  ( -8 p_1 p_2 ) \, 
  \frac{1}{8} {\rm ln}\left( \frac{p^2}{\mu^2} \right) 
   \, . 
 } \label{eq:02_c} 
\end{equation} 
 Hence, eqs.~(\ref{eq:02_b}) and (\ref{eq:02_c}) give 
\begin{equation} 
 \Pi_{02}(p) = 0 \, . 
\end{equation} 

 The two-point function $\Pi_{2 -}(p)$ 
receives corrections from the diagrams of topology 
in Fig.~\ref{fig:diagram}(c), 
\begin{eqnarray} 
 \displaystyle{ 
  \left. 
   \Pi_{2 -}(p) 
  \right|_{Y^I} 
 } 
 &=& 
 \displaystyle{ 
  \frac{1}{4\pi} 
  (-\frac{D-1}{3}) 
  \frac{(p_1)^3 p_2}{p^2} \, , 
 } \nonumber \\ 
 \displaystyle{ 
  \left. 
   \Pi_{2 -}(p) 
  \right|_{A_r} 
 } 
 &=& 
 \displaystyle{ 
  \frac{1}{4\pi} 
  \left(- \frac{1}{6} \right) 
  \frac{(p_1)^3 p_2}{p^2} 
   \, , 
 } \nonumber \\ 
 \displaystyle{ 
  \left. 
   \Pi_{2 -}(p) 
  \right|_{\rm G} 
 } 
 &=& 
 \displaystyle{ 
  \frac{1}{4\pi} \, 
   \frac{53}{6} \frac{(p_1)^3 p_2}{p^2} 
    \, , 
 } 
\end{eqnarray} 
which give in total 
\begin{equation} 
 \displaystyle{ 
  \Pi_{2 -}(p) 
 } 
 = 
 \displaystyle{ 
  \frac{1}{4\pi} \, 
   \frac{27 - D}{3} \frac{(p_1)^3 p_2}{p^2} 
    \, . 
 } 
\end{equation} 

 Two types of diagrams shown in 
Fig.~\ref{fig:diagram}(b) and (c) contributes 
to $\Pi_{0 -}(p)$. 
 The one propagator contribution 
$\left. \Pi_{0 -}(p) \right|^{(b)}$ is found as 
\begin{eqnarray} 
 \displaystyle{ 
  \left. 
   \Pi_{0 -}(p) 
  \right|_{Y^I}^{(b)} 
 } 
 &=& 
 0 \, , 
 \nonumber \\ 
 \displaystyle{ 
  \left. 
   \Pi_{0 -}(p) 
  \right|_{A_r}^{(b)} 
 } 
 &=& 
 \displaystyle{ 
  \frac{1}{4\pi} 
  \left( - p^2 + 2 (p_2)^2 \right) 
  \frac{1}{8} {\rm ln} \left( \frac{p^2}{\mu^2} \right) 
    \, , 
 } \nonumber \\ 
 \displaystyle{ 
  \left. 
   \Pi_{0 -}(p) 
  \right|_{\rm G}^{(b)} 
 } 
 &=& 
 \displaystyle{ 
  \frac{1}{4\pi} 
  \left( - 5 p^2 + 6 (p_2)^2 \right) 
  \frac{1}{8} {\rm ln}\left( \frac{p^2}{\mu^2} \right) 
   \, , 
 } 
\end{eqnarray} 
which give 
\begin{equation} 
 \displaystyle{ 
  \left. 
   \Pi_{0 -}(p) 
  \right|^{(b)} 
  = 
  \frac{1}{4\pi} 
  \left( - 6 p^2 + 8 (p_2)^2 \right) 
  \frac{1}{8} {\rm ln}\left( \frac{p^2}{\mu^2} \right) 
   \, . 
 } \label{eq:0-_b} 
\end{equation} 
 The two-particle irreducible contribution 
$\left. \Pi_{0 -}(p) \right|^{(c)}$ 
is found as 
\begin{eqnarray} 
 \displaystyle{ 
  \left. 
   \Pi_{0 -}(p) 
  \right|_{Y^I}^{(c)} 
 } 
 &=& 
 0 \, , 
  \nonumber \\ 
 \displaystyle{ 
  \left. 
   \Pi_{0 -}(p) 
  \right|_{A_r}^{(c)} 
 } 
 &=& 
 0 \, , 
  \nonumber \\ 
 \displaystyle{ 
  \left. 
   \Pi_{0 -}(p) 
  \right|_{\rm G}^{(c)} 
 } 
 &=& 
 \displaystyle{ 
  \frac{1}{4\pi} 
  \left( 6 p^2 - 8 (p_2)^2 \right) 
  \frac{1}{8} {\rm ln}\left( \frac{p^2}{\mu^2} \right) 
   \, , 
 } 
\end{eqnarray} 
the sum of which becomes 
\begin{equation} 
 \displaystyle{ 
  \left. 
   \Pi_{0 -}(p) 
  \right|^{(c)} 
  = 
  \frac{1}{4\pi} 
  \left( 6 p^2 - 8 (p_2)^2 \right) 
  \frac{1}{8} {\rm ln}\left( \frac{p^2}{\mu^2} \right) 
   \, . 
 } \label{eq:0-_c} 
\end{equation} 
 Hence, 
the sum of eqs.~(\ref{eq:0-_b}) and (\ref{eq:0-_c}) vanishes; 
\begin{equation} 
 \Pi_{0 -}(p) = 
  0 \, . 
\end{equation} 
 
\end{document}